\documentclass[aps, prd, amsmath, floats, floatfix, twocolumn,superscriptaddress, nofootinbib, showpacs]{revtex4}

\usepackage{graphicx}
\usepackage{amsmath,amssymb}
\usepackage{amsfonts}
\usepackage{xspace} 
\usepackage[usenames]{color}
\usepackage{dcolumn}
\usepackage{bm}

\newcommand{\Caltech}{\affiliation{Theoretical Astrophysics 350-17,
    California Institute of Technology, Pasadena, CA 91125}}
\newcommand{\Cornell}{\affiliation{Center for Radiophysics and Space
    Research, Cornell University, Ithaca, New York, 14853}}
\newcommand{\Maryland}{\affiliation{Maryland Center for Fundamental
    Physics, Department of Physics, University of Maryland, College
    Park, MD 20742}}
\newcommand{\CITA}{\affiliation{Canadian Institute for Theoretical Astrophysics,
    University~of~Toronto, Toronto, Ontario M5S 3H8, Canada}}

\newcommand{\vrr}{\mbox{\boldmath${r}$}}
\newcommand{\vR}{\mbox{\boldmath${R}$}}

\newcommand{\vp}{\mbox{\boldmath${p}$}}
\newcommand{\vP}{\mbox{\boldmath${P}$}}
\newcommand{\vpstar}{\mbox{\boldmath${p^*}$}}
\newcommand{\vPstar}{\mbox{\boldmath${P^*}$}}
\newcommand{\vSa}{\mbox{\boldmath${S_1}$}}
\newcommand{\vSb}{\mbox{\boldmath${S_2}$}}
\newcommand{\vS}{\mbox{\boldmath${S}$}}
\newcommand{\vs}{\mbox{\boldmath${s}$}}
\newcommand{\vL}{\mbox{\boldmath${L}$}}
\newcommand{\vF}{\mbox{\boldmath${F}$}}
\newcommand{\vsigma}{\mbox{\boldmath${\sigma}$}}

\begin{document}

\title{Effective-one-body waveforms calibrated to numerical relativity simulations: 
coalescence of non-precessing, spinning, equal-mass black holes}

\author{Yi Pan} \Maryland %
\author{Alessandra Buonanno} \Maryland %
\author{Luisa T. Buchman} \Caltech %
\author{Tony Chu} \Caltech %
\author{Lawrence E. Kidder} \Cornell %
\author{Harald P. Pfeiffer} \CITA \Caltech  %
\author{Mark A. Scheel} \Caltech %

\begin{abstract}
  We present the first attempt at calibrating the effective-one-body
  (EOB) model to accurate numerical-relativity simulations of
  spinning, non-precessing black-hole binaries. Aligning the EOB and
  numerical waveforms at low frequency over a time interval of $1000
  M$, we first estimate the phase and amplitude errors in the
  numerical waveforms and then minimize the difference between
  numerical and EOB waveforms by calibrating a handful of
  EOB-adjustable parameters. In the equal-mass, spin aligned case, we
  find that phase and fractional amplitude differences between the
  numerical and EOB (2,2) mode can be reduced to 0.01 radians and 1\%,
  respectively, over the entire inspiral waveforms. In the equal-mass,
  spin anti-aligned case, these differences can be reduced to 0.13
  radians and 1\% during inspiral and plunge, and to 0.4 radians and
  10\% during merger and ringdown. The waveform agreement is within
  numerical errors in the spin aligned case while slightly over
  numerical errors in the spin anti-aligned case. Using Enhanced LIGO
  and Advanced LIGO noise curves, we find that the overlap between the
  EOB and the numerical (2,2) mode, maximized over the initial phase
  and time of arrival, is larger than 0.999 for binaries with total
  mass $30\mbox{--} 200 M_\odot$. In addition to the leading (2,2)
  mode, we compare four subleading modes.  We find good amplitude and
  frequency agreements between the EOB and numerical modes for both
  spin configurations considered, except for the (3,2) mode in the
  spin anti-aligned case.  We believe that the larger difference in
  the (3,2) mode is due to the lack of knowledge of post-Newtonian
  spin effects in the higher modes.
 \end{abstract}

\date{\today \hspace{0.2truecm}}


\maketitle

\section{Introduction}
\label{sec:intro}

Coalescing black-hole binaries are among the most promising sources
for the current and future laser-interferometer gravitational-wave
detectors such as LIGO/Virgo\cite{Barish:1999,Waldman:2006,
  Acernese-etal:2006} and LISA~\cite{schutz_lisa_science}. 

In general relativity, black holes are defined only by their masses
and spins; thus generically, a black-hole binary depends on eight
parameters $(m_1,\boldsymbol{S}_1, m_2, \boldsymbol{S}_2)$. Hence,
when black holes carry spins, it is expected that tens of thousands of
waveform templates may be needed in order to extract the gravitational-wave
signal from the noise using matched-filtering techniques.  Considering
the high computational cost of running numerical-relativity
simulations of spinning binary black holes (on the order of $75,000$
CPU hours for $15$ orbits, moderate spins and mild mass ratios) and
the large binary parameter space, it will be impractical for numerical
relativity alone to provide data analysts with a template bank. The
work at the interface between analytical and numerical relativity
\cite{Buonanno-Cook-Pretorius:2007,Pan2007,Ajith-Babak-Chen-etal:2007,Ajith-Babak-Chen-etal:2007b,
  Buonanno2007,Damour2007a,DN2007b,DN2008,Boyle2008a,Damour2009a,Buonanno:2009qa}
has demonstrated the possibility of modeling analytically the dynamics
and the gravitational-wave emission of coalescing {\it nonspinning}
black holes, thus providing data analysts with preliminary analytical
template families to be used for the searches. The next important step
is to extend those studies to {\it spinning, precessing} black
holes. The present paper represents the first attempt in
this direction, although limited to {\em non-precessing} waveforms, 
within the effective-one-body (EOB) formalism~\cite{Buonanno99,Buonanno00,2000PhRvD..62h4011D} 
of spinning black holes~\cite{Damour01c,DJS08}.  Recently, Ref.~\cite{Ajith:2009bn}
constructed a template family of spinning, non-precessing black-hole
binaries using a phenomenological approach, where the
numerical-relativity waveforms are fitted to templates which resemble
the post-Newtonian (PN) expansion, but in which the coefficients
predicted by PN theory are replaced by many arbitrary
coefficients calibrated to numerical simulations.

The first EOB Hamiltonian that included spin effects was worked out in
Ref.~\cite{Damour01c}.  In Ref.~\cite{Buonanno06}, the authors used
the non-spinning EOB Hamiltonian augmented with PN spin terms to carry
out the first exploratory study of the dynamics and gravitational
radiation of spinning black-hole binaries during inspiral, merger and
ringdown.  Subsequently, Ref.~\cite{DJS08} extended the EOB
Hamiltonian of Ref.~\cite{Damour01c} to include next-to-leading order
spin-orbit couplings.  In those descriptions, the effective particle
is endowed not only with a mass $\mu$, but also with a spin
$\vsigma$. As a consequence, the effective particle interacts with the
effective Kerr background (having spin $\vS_{\rm Kerr}$) both via a
geodesic-type interaction and via an additional spin-dependent
interaction proportional to its spin $\vsigma$. The EOB Hamiltonian
developed in Refs.~\cite{Damour01c,DJS08} (with nonspinning PN
couplings through 3PN order) clarified several features of spinning
two-body dynamics. However, as we shall discuss below, it is not
straightforward to extend this Hamiltonian to include higher-order
nonspinning PN couplings, such as the 4PN or 5PN adjustable parameters
recently calibrated to numerical-relativity
simulations~\cite{Damour2009a,Buonanno:2009qa}.  Moreover, the EOB
Hamiltonian of Ref.~\cite{DJS08}, based on an {\em ad hoc}
test-particle limit, does not reduce to the Hamiltonian of a spinning
test particle in Kerr spacetime. More recently, Ref.~\cite{BRB}
derived the canonical Hamiltonian of a spinning test particle in a
generic curved spacetime at linear order in the particle spin.  The
construction of an improved EOB Hamiltonian based on the results of
Ref.~\cite{BRB} is currently under development. Despite the
limitations mentioned above, the EOB Hamiltonian of Ref.~\cite{DJS08}
is an excellent starting point for exploring the calibration of
numerical-relativity waveforms of spinning black holes within the EOB
formalism. Thus, we have used it in this first exploratory study,
augmenting it with a few {\it adjustable} parameters that we shall
calibrate to two numerical-relativity simulations.  For the EOB
non-conservative dynamics, we use the gravitational-wave energy flux
which includes spin effects and which has been computed using the
factorized multipolar waveforms of Refs.~\cite{Pan:2009,DIN}.

The two numerical-relativity simulations we shall use describe the
evolution of equal-mass, equal-spin, non-precessing black-hole
binaries. They are produced by the pseudospectral code SpEC of the
Caltech-Cornell-CITA collaboration. In these two configurations, the
spins are either aligned (``up-up'', or UU) or anti-aligned
(``down-down'', or DD) with the orbital angular momentum, and have
dimensionless magnitude $\chi_1=\chi_2=0.43655$ for the UU
configuration, and $0.43757$ for the DD configuration. The UU
simulation lasts for about 28 gravitational-wave cycles or until
$t=3250M$, and stops about three gravitational-wave cycles before
merger. The DD simulation lasts for about 22 gravitational-wave cycles
or until $t = 2500M$, and contains the full inspiral, merger, and
ringdown waveform.  
Detailed information on the numerical simulation
of the DD configuration can be found in Ref.~\cite{Chu:2009md}. 

For zero spin, the EOB-model considered here agrees with
the waveform of the equal-mass non-spinning binary black
hole~\cite{Scheel2008} to a similar degree as the model 
constructed in our earlier work~\cite{Buonanno:2009qa}.  It differs from the model
presented in~\cite{Buonanno:2009qa} by its modeling of the energy 
flux of gravitational-wave radiation.

This paper is organized as follows. In Sec.~\ref{sec:EOB}, we describe
the spin EOB model adopted in this paper, including its adjustable
parameters.  In Sec.~\ref{sec:EOBcalibration}, we calibrate the spin
EOB model to the numerical simulations, and discuss the impact of our
results on data analysis. Finally, Sec.~\ref{sec:conclusions}
summarizes our main conclusions.

\section{EOB model for spinning black-hole binaries}
\label{sec:EOB}

In this section, we describe the spin EOB model adopted in our study
and its adjustable parameters. Henceforth we use natural units
$G=c=1$.  We use $m_i$, $\boldsymbol{X}_i$, $\boldsymbol{P}_i$, and
$\vS_i$ to denote the mass, the position vector, the linear momentum
vector, and the spin vector of the $i$-th body.  We work in the
center-of-mass frame defined by
$\boldsymbol{P}_1+\boldsymbol{P}_2=0$. The two body system is
described by the relative position $\vR\equiv\boldsymbol{X}=
\boldsymbol{X}_1-\boldsymbol{X}_2$ and the relative linear momentum
$\vP\equiv \boldsymbol{P}_1=-\boldsymbol{P}_2$. For convenience, we
define reduced variables
\begin{equation}
\vrr\equiv\frac{\vR}{M} \qquad \vp\equiv\frac{\vP}{\mu} \,,
\end{equation}
where $M\equiv m_1+m_2$ and $\mu\equiv m_1m_2/(m_1+m_2)$.

\subsection{EOB conservative dynamics}
\label{sec:EOBdynamics}

Following Refs.~\cite{Damour01c,DJS08}, we assume that the effective
particle in the EOB description is endowed not only with a mass
$\mu$, but also with a spin $\vsigma$. As a
consequence, the effective particle interacts with the effective Kerr
background (having spin $\vS_{\rm Kerr}$ and mass $M$) both
via a geodesic-type interaction and via an additional spin-dependent
interaction proportional to its spin $\vsigma$. We define the
Kerr-like parameter $a$ as $a\equiv S_\mathrm{Kerr}/M$, where
$S_\mathrm{Kerr}$ denotes the modulus of the deformed-Kerr spin vector
$\boldsymbol{S}_\mathrm{Kerr}$. 
Following Ref.~\cite{Damour01c}, we write the
effective Kerr contravariant metric components in a fixed
Cartesian-like coordinate system.  This is done by introducing
\begin{align}
n^i &\equiv \frac{X^i}{R}, \quad
s^i \equiv \frac{S^i_\mathrm{Kerr}}{S_\mathrm{Kerr}}, \quad
\cos\theta \equiv n^i s^j \delta_{ij},
\nonumber\\
\rho &\equiv \sqrt{R^2 + {a^2} \cos^2\theta},
\end{align}
and 
\begin{equation}
\alpha \equiv (-g^{00}_\mathrm{eff})^{-1/2}, \quad
\beta^i \equiv \frac{g^{0i}_\mathrm{eff}}{g^{00}_\mathrm{eff}},
\quad \gamma^{ij} \equiv g^{ij}_\mathrm{eff}
- \frac{g^{0i}_\mathrm{eff}\,g^{0j}_\mathrm{eff}}{g^{00}_\mathrm{eff}},
\label{abg}
\end{equation}
and writing the contravariant metric components as
\begin{subequations}
\label{cometric2}
\begin{align}
g^{00}_\mathrm{eff}
&= -\frac{(R^2+a^2)^2-a^2\,\Delta_t(R)\,\sin^2\theta}
{\rho^2\,\Delta_t(R)},
\\
g^{0i}_\mathrm{eff}
& = -\frac{a\,(R^2+a^2-\Delta_t(R))}{\rho^2\,\Delta_t(R)}
(\vs\times\vR)^i,\\
g^{ij}_\mathrm{eff} &= \frac{1}{\rho^2}\Big[\Delta_R(R)\,n^i\,n^j
+ R^2\,(\delta^{ij}-n^in^j)\Big]
\nonumber\\[1ex]&\quad
- \frac{a^2}{\rho^2\,\Delta_t(R)}
(\vs\times\vR)^i(\vs\times\vR)^j,
\end{align}
\end{subequations}
where~\footnote{We denote with $P_m^n$ the operation of taking the $(n,m)$-Pad\'e approximant.} 
\begin{subequations}
\begin{eqnarray}
\label{Deltat}
\Delta_t(R) &=&  R^2\,P_m^n\left [{A}(R) + \frac{a^2}{R^2}\right ]\,,\\
\Delta_R(R) &=& \frac{\Delta_t(R)}{D(R)}\,. 
\label{DeltaR}
\end{eqnarray}
\end{subequations}
The Taylor approximants to the coefficients $A(R)$ and $D(R)$ can be
written as
\begin{subequations}\begin{eqnarray}
  \label{coeffA}
  A_{k}(r) &=& \sum_{i=0}^{k+1} \frac{a_i(\nu)}{r^i} \,,\\
\label{coeffD}
  D_{k}(r) &=& \sum_{i=0}^k \frac{d_i(\nu)}{r^i}\,.
\end{eqnarray}
\end{subequations}
The functions $A_{k}(r)$ and $D_{k}(r)$ all depend on the symmetric
mass ratio $\nu\equiv\mu/M$ through the $\nu$--dependent coefficients
$a_i(\nu)$ and $d_i(\nu)$.  These coefficients are currently known
through 3PN order (i.e. up to $k=4$) and can be read off from
Eqs.~(47) and (48) in Ref.~\cite{Boyle2008a}.  It is worth noticing
that although through 3PN order the Pad\'e approximant to the function
$\Delta_t(R)$ of Eq.~\eqref{Deltat} does not pose any
problem~\cite{DJS08}, when including 4PN and 5PN order coefficients,
the Pad\'e approximant develops poles for several spin values $a$. In
particular, poles are present at large separation when $a> 0.7 M$ 
and the 4PN and 5PN order coefficient $a_5$ and
$a_6$ are included.~\footnote{Poles also develop when only the 4PN order 
coefficient $a_5$ is included and $a> 0.96 M$.}  Those poles could be regularized by adding in
$A_{k}(r)$ higher-order spin terms $a^2\,\tilde{a}_3(\nu)/r^5$,
$a^2\,\tilde{a}_4(\nu)/r^6$ and choosing for the coefficients
$\tilde{a}_3(1/4)$ and $\tilde{a}_4(1/4)$ negative large values ($\sim
-100$). Since in this first exploratory study we investigate only
numerical simulations of moderate spins, we do not include any
regularization of the poles, and consider only the 4PN order
coefficient $a_5$. In the nonspinning case~ \cite{Buonanno2007,
  Damour2007a, Damour2009a, Buonanno:2009qa}, the coefficient $a_5$
plays an important role in improving the agreement between the EOB and
numerical waveforms.  Here, we choose for $a_5$ the value obtained by
taking the nonspinning limit of the spin EOB model and calibrating it
to the equal-mass black-hole waveform of~\cite{Scheel2008},
following~\cite{Buonanno:2009qa}.  In this way, we obtain $a_5(1/4)=
1.775$; thus, $a_5$ is no longer an adjustable parameter in the spin
EOB model.

In Eq.~\eqref{Deltat}, we choose $m=1$ and $n=4$ so that
$\Delta_t(R)/R^2$ in the limit of $a \rightarrow 0$ reduces to the
nonspinning $A(R)$ used in Refs.~\cite{Buonanno2007, Damour2007a,
  Damour2009a, Buonanno:2009qa}, and we choose the same 3PN $D(R)$
function used in those references. Therefore, in the spin EOB model,
we have
\begin{eqnarray}
\frac{\Delta_t(R)}{R^2}&=& \frac{{\rm Num}(\Delta_t)}{{\rm Den}(\Delta_t)}\,, \\
D(r)&=& \frac{r^3}{r^3+6\,\nu r+2\,\nu(26-3\,\nu)}\,,
\end{eqnarray}
with
\begin{eqnarray}
  {\rm Num}(\Delta_t) &=& r^3\,[32 -24 \nu - 4 a_4(\nu) - a_5(\nu) \nonumber \\ 
&& \quad- (32-4\nu)\chi^2+6\chi^4] \nonumber \\ 
&& + r^4 [a_4(\nu)  - 16 +8 \nu + 12 \chi^2 - \chi^4]\,,
\end{eqnarray} 
and
\begin{widetext}
\begin{eqnarray} 
{\rm Den}(\Delta_t) &=& -a_4^2(\nu) - 8a_5(\nu) - 8
  a_4(\nu) \nu + 2 a_5(\nu) \nu - 16 \nu^2 + (4a_5(\nu)-8a_4(\nu)-8\nu^2)\chi^2 \nonumber\\
&& + (2 a_4(\nu) - 12\nu)\chi^4 - \chi^8 + r\,[-8 a_4(\nu) -4 a_5(\nu) -2 a_4(\nu)  \nu -  16 \nu^2 \nonumber \\
&& + (a_5(\nu)-16\nu)\chi^2-2\nu \chi^4-2\chi^6] + r^2\,[-4a_4(\nu) - 2a_5(\nu) -16 \nu \nonumber \\
&& - a_4(\nu) \chi^2 - 4\chi^4+\chi^6] + r^3\,[-2 a_4(\nu) - a_5(\nu) - 8 \nu - (8-4\nu)\chi^2 + 4\chi^4] \nonumber \\
&& + r^4\, [-16 + a_4(\nu) + 8 \nu + 12 \chi^2 - \chi^4]\,,
\end{eqnarray}
\end{widetext}
where $\chi\equiv a/M$ and $a_4(\nu) = \left ({94}/{3}-{41}/{32}\pi^2 \right )\,\nu$. 
Making use of Eqs.\ \eqref{abg} and \eqref{cometric2}, we can derive
\begin{subequations}
\begin{align}
\alpha &= \rho \sqrt{\frac{\Delta_t(R)}{(R^2+a^2)^2-a^2\,\Delta_t(R)\,\sin^2\theta}},\\
\beta^i &=
\frac{a\,(R^2+a^2-\Delta_t(R))}{(R^2+a^2)^2-a^2\,\Delta_t(R)\,\sin^2\theta}
(\vs\times\vR)^i, \\
\gamma^{ij} &= g^{ij}_\mathrm{eff} + \frac{\beta^i\,\beta^j}{\alpha^2}.
\end{align}
\end{subequations}
The EOB effective Hamiltonian reads~\cite{Damour01c,DJS08}
\begin{eqnarray}
\label{Heff}
H_\mathrm{eff}(\vR,\vP,\vS_1,\vS_2) &=& 
H_{\rm eff\, Kerr}(\vR,\vP,\vS_{\rm Kerr}) \nonumber \\
&+& H_{\rm eff\,part}(\vR,\vP,\vsigma)\,,
\end{eqnarray}
and 
\begin{eqnarray}
\label{Heff1}
&& H_{\rm eff\,Kerr}(\vR,\vP,\vS_{\rm Kerr}) 
= \beta^i P_i + \alpha\,\sqrt{\mu^2  + \gamma^{ij} P_i P_j + {Q_4}}\,, \nonumber 
\\
\label{Heff2}
&& H_{\rm eff\,part}(\vR,\vP,\vsigma) = 
\frac{R^2+a^2-\Delta_t(R)}{(R^2+a^2)^2-a^2\,\Delta_t(R)\,\sin^2\theta}\,\vL\cdot\vsigma \nonumber\\
\end{eqnarray}
where ${Q}_4(P_i)$ is a quartic-momentum term at 3PN order independent
of spins~\cite{2000PhRvD..62h4011D} and $\vL\equiv\vR\times\vP$ is the
orbital angular momentum. In this paper, as a first attempt, we use
the same spin coupling for the spin $\vsigma$ suggested in
Ref.~\cite{DJS08}, even though it does not reduce to the spinning
test-particle limit~\cite{BRB} at PN orders higher than 2.5PN.

In order for $H_{\rm eff}$ to match the PN-expanded spin-orbit Hamiltonian through 
2.5PN order, we need to require that the sum of the spin-orbit couplings of $H_{\rm eff\, Kerr}$
and $H_{\rm eff\, part}$ gives 
\begin{equation}
\label{constraint}
\left [H_{\rm eff\, Kerr}+ H_{\rm eff\, part}\right ]_{\rm SO} \simeq \frac{2}{R^3} \mathbf{L} \cdot 
\Big(\frac{1}{2} g^\mathrm{eff}_{S} \vS
+ \frac{1}{2} g^\mathrm{eff}_{S^*} \vS^* \Big),
\end{equation}
where 
\begin{subequations}
\begin{align}
\vS &\equiv {\bf S}_1 + {\bf S}_2\,,\\
\vS_* &\equiv \frac{m_2}{m_1}{\bf S}_1
+ \frac{m_1}{m_2}{\bf S}_2\,,
\end{align}
\end{subequations}
and where the two effective gyro-gravitomagnetic ratios $g^\mathrm{eff}_S$ and $g^\mathrm{eff}_{S^*}$ read~\cite{DJS08}
\begin{subequations}\begin{eqnarray}
\label{anu}
g^\mathrm{eff}_S &\equiv&  2 + \left [\frac{3}{8}\nu+a(\nu)\right ] {\bf p}^2 \nonumber \\
&& - \left [\frac{9}{2}\nu+3a(\nu)\right ]({\bf n}\cdot{\bf p})^2 \nonumber \\
&&  - \left [\nu+a(\nu)\right ]\frac{1}{r}\,, \\
\label{bnu}
g^\mathrm{eff}_{S^*} &\equiv& \frac{3}{2} + \left [-\frac{5}{8}+\frac{1}{2}\nu+b(\nu)\right ]{\bf p}^2 \nonumber \\
&& - \left [\frac{15}{4}\nu+3b(\nu)\right ]({\bf n}\cdot{\bf p})^2 \nonumber \\
&& - \left [\frac{1}{2}+\frac{5}{4}\nu+b(\nu)\right ]\frac{1}{r}.
\end{eqnarray}
\end{subequations}
Here $a(\nu)$ and $b(\nu)$ are two {\it gauge} parameters related to the
freedom of applying a canonical transformation involving spin
variables. If we knew the exact Hamiltonian, the choice of
  these parameters should not affect the physics of the EOB
  model. However, since we start with an approximate Hamiltonian that
  reproduces the spin-orbit couplings only through 2.5PN order, we expect the EOB
  model to depend on the choice of $a(\nu)$ and $b(\nu)$. Considering the structure 
  of the gyro-gravitomagnetic ratios, such dependence
  should start at 3.5PN order as a spin-orbit coupling term. Because
of this dependence, $a(\nu)$ and $b(\nu)$ can be used as adjustable
parameters. 

Moreover, in order for $H_{\rm eff}$ to match the PN-expanded spin-spin Hamiltonian through
2PN order, the simplest choice is to require that the Kerr spin~\cite{DJS08}
\begin{equation}
\vS_{\rm Kerr} = \vS + \vS_*\,.
\end{equation}
As a consequence, Eq.~(\ref{constraint}) implies 
\begin{equation}
\mathbf{\sigma} = \frac{1}{2} (g^\mathrm{eff}_{S}-2) \vS
+ \frac{1}{2} (g^\mathrm{eff}_{S^*}-2) \vS^*\,.
\end{equation}
To include higher-order spin-spin contributions in the EOB effective Hamiltonian, we introduce a 
3PN spin-spin term whose coefficient $a^{\rm 3PN}_{\rm SS}$ is currently unknown and can be used as an adjustable parameter
\begin{eqnarray}
\label{HeffSS3PN}
H_\mathrm{eff}(\vR,\vP,\vS_1,\vS_2) &=& 
H_{\rm eff\, Kerr}(\vR,\vP,\vS_{\rm Kerr}) \nonumber \\
&+& H_{\rm eff\,part}(\vR,\vP,\vsigma) \nonumber\\
&+& a^{\rm 3PN}_{\rm SS}\nu\,\frac{\vS_{\rm Kerr}\cdot\vS^*}{R^4}\,.
\end{eqnarray}
Finally, the EOB Hamiltonian is 
\begin{equation}
\label{hreal}
H_{\rm real} = Mc^2
\sqrt{1+2\nu\Big(\frac{H_{\rm eff}}{\mu c^2}-1\Big)}.
\end{equation}
In summary, in this first exploratory study, we choose to employ only
two adjustable parameters\footnote{We find that $b(\nu)$ is
strongly degenerate with $a(\nu)$.} : $b(\nu)$
which introduces a spin-orbit
term at 3.5PN order, and $a^{\rm 3PN}_{\rm SS}$ which
introduces a 3PN spin-spin term. As we shall see, these two
adjustable parameters are sufficient to reduce the phase and amplitude
differences between EOB and numerical waveforms of the UU and DD
configurations to (almost) the numerical error. The remaining
flexibility of the spin EOB model can be exploited in the future when
numerical relativity simulations of other spin configurations will
become available.  Thus, for the rest of the paper, we set 
$a(\nu)$ in Eq.~\eqref{anu} to zero.

Within the Hamiltonian approach, radiation-reaction effects can be incorporated into 
the dynamics in the following way~\cite{Buonanno00,Buonanno06}:
\begin{eqnarray}\label{eompr1}
\frac{dX^i}{dt} &=& \{ X^i , H_{\rm real} \} = \frac{\partial H_{\rm real}}{\partial P_i} \,, \\
\label{eompr2}
\frac{dP_i}{dt} &=& \{ P_i , H_{\rm real} \} + F_i = - \frac{\partial H_{\rm real}}{\partial X^i} + F_i \, .
\end{eqnarray}
Here, $F_i$ denotes the non-conservative force, which is added to the
evolution equation of the (relative) momentum to take into account
radiation-reaction effects. Following Ref.~\cite{Buonanno06}, we use
\footnote{We notice that this choice of the radiation-reaction force 
  introduces a radial component of the force
  $\vR\cdot\vF\propto\vR\cdot\vP=R\,P_R$. In the nonspinning EOB
  models, this component is usually ignored~\cite{Damour2009a,
    Buonanno:2009qa}.}
\begin{equation}
F_i = \frac{1}{\Omega \, \vert \vL \vert} \, \frac{dE}{dt} \, P_i \,,
\end{equation}
where $\Omega$ is the orbital frequency and $\vL$ is the orbital angular momentum. The gravitational-wave energy 
flux $dE/dt$ is obtained by summing over the gravitational-wave modes $(l,m)$ as
\begin{equation}
\frac{dE}{dt}=\frac{1}{16\pi}\sum_{\ell=2}^8\sum_{m=-\ell}^{\ell}\left|\dot{h}_{\ell m}\right|^2\,,
\end{equation}
which reduces to the following expression for circular equatorial orbits in the adiabatic approximation:
\begin{equation}
\frac{dE}{dt}=\frac{1}{16\pi}\sum_{\ell=2}^8\sum_{m=-\ell}^{\ell}(m\,\hat{\Omega})^2\left|h_{\ell m}\right|^2\,,
\end{equation}
where $\hat{\Omega}$ is the reduced orbital frequency
$\hat{\Omega}\equiv M\Omega$.  We shall define the EOB waveforms
$h_{\ell m}$ in Sec.~\ref{sec:EOBinspiralwaveforms}.  The equations of
motion for the spins are simply obtained through the equations
\begin{eqnarray}
\frac{d }{dt} \vS_1 = \{\vS_1,H_{\rm real} \} = \frac{\partial H_{\rm real}}{\partial \vS_1} \times \vS_1 \,,\\
\frac{d }{dt} \vS_2 = \{\vS_2,H_{\rm real} \} = \frac{\partial H_{\rm real}}{\partial \vS_2} \times \vS_2 \,.
\end{eqnarray}
In the nonspinning case, it is
useful~\cite{Damour2007,Buonanno:2009qa} to replace the radial
momentum $P_R$ with $P_{R^*}$, the conjugate momentum of the EOB {\it
  tortoise} radial coordinate $R^*$: $dR^*/dR=\sqrt{D}/A$.  This
replacement improves the numerical stability of the EOB equations of
motion because $P_R$ diverges when approaching the zero of $A(r)$ (the
EOB event horizon) but $P_{R^*}$ does not.  Therefore, in the spinning
EOB Hamiltonian, we similarly choose to use the conjugate momentum to
the tortoise radial coordinate of the $\nu$-deformed Kerr geometry:
\begin{equation}\label{tortoise}
\frac{dR^*}{dR}=\frac{R^2+a^2}{\sqrt{\Delta_t\,\Delta_R}}\equiv \frac{1}{\xi_a(R)}\,.
\end{equation}
In the limit $a\rightarrow 0$, Eq.~\eqref{tortoise} reduces to the
nonspinning EOB tortoise coordinate defined above. In the
limit $\nu\rightarrow 0$, Eq.~\eqref{tortoise} reduces to the tortoise
coordinate of the Kerr geometry: $dR^*/dR=(R^2+a^2)/\Delta$. Since the EOB
Hamiltonian and Hamilton equations are written in Cartesian
coordinates, some algebra is needed to rewrite them to include this 
transform of the radial coordinate. In Appendix~\ref{sec:tortoise}, we
write down explicitly the transformed EOB Hamiltonian and Hamilton
equations in Cartesian coordinates. In particular, Eqs.~\eqref{eompr1}
and \eqref{eompr2} should be replaced by Eqs.~\eqref{eomprstar1} 
and~\eqref{eomprstar2}.

Initial conditions for the Hamilton equations are constructed following 
the prescription of Ref.~\cite{Buonanno06}, which provided post-circular 
initial data for quasi-spherical orbits when neglecting spin-spin and next-to-leading order 
spin-orbit effects. Note that exact circular orbits cease to exist in the
conservative dynamics when spin-spin and next-to-leading order
spin-orbit effects are present, except for special 
configurations in which the spins are aligned or antialigned with the orbital angular momentum. 
We start each evolution at a large initial separation of $50M$. The EOB 
trajectory is sufficiently circularized when reaching a separation of $\sim 16M$, 
where numerical waveforms start. In this way, we remove the residual eccentricity in 
the EOB trajectory due to imperfect initial conditions, while physical 
eccentricity due to spin effects is preserved.

As a final remark, the spin variable in the EOB model is 
the {\it constant} spin variable, i.e., its magnitude does not change 
during precession~\cite{Blanchet-Buonanno-Faye:2006}. We identify 
it with the spin variable in the numerical simulation, which also 
remains constant during the evolution~\cite{Chu:2009md}.

\subsection{Characteristics of EOB orbits for spinning, non-precessing black holes}
\label{sec:EOBstudydynamics}

Here we investigate certain properties of the spin EOB Hamiltonian 
that are crucial when building the complete EOB
model. Specifically, we check the existence and behavior of the
innermost stable circular orbit (ISCO), the light ring (photon orbit)
and the maximum of the EOB orbital frequency. This study is restricted to
circular equatorial orbits in the spin aligned or anti-aligned
cases. For convenience, we consider the EOB  Hamiltonian written 
in spherical coordinates; we fix $\theta=\pi/2$ and set the
conjugate momentum $P_\theta=0$.

The ISCO position is obtained by solving $ \partial H(R,P_{R^*}=0,P_\phi)/{\partial R}=0, 
\partial^2 H(R,P_{R^*}=0,P_\phi)/{\partial R^2}=0$
where $P_{R^*}$ and $P_\phi$ are conjugate momenta of the tortoise
radial coordinate $R^*$ and the orbital phase $\phi$, respectively.
In the nonspinning limit, we find the following $\nu$-correction of
the ISCO frequency relative to the Schwarzschild value
\begin{eqnarray}
\hat{\Omega}_{\rm ISCO}&=&6^{-3/2}\times \nonumber\\
&&\left[1+0.9837\nu+1.2543\nu^2+5.018\nu^3+\mathcal{O}(\nu^4)\right]\,,\nonumber\\
\end{eqnarray}
where $\mathcal{O}(\nu^4)$ terms contribute less than $1\%$ to the
correction. In the test-particle limit, the coefficient of the linear
$\nu$-correction term, $0.9837$, should be compared to the recently
available self-force result \cite{barack:2009} (transformed to the gauge 
condition and mass convention used in the EOB formalism
by Ref.~\cite{Damour:2009sm}) of $1.2513$. The relative difference of $21\%$
is due to the fact that our nonspinning EOB Hamiltonian, although
calibrated to equal-mass numerical simulations, does not capture all
the $\nu$-dependence correctly at 4PN order.~\footnote{We notice that if we used the 
4PN and 5PN coefficients, $a_5$ and $a_6$, suggested in 
Ref.~\cite{Damour:2009sm}, we would obtain poles in the 
function $\Delta_t(R)$ for $|a|>0.75M$. Moreover, if we adopted 
the values of $a_5$ and $a_6$, suggested in 
Ref.~\cite{Damour:2009sm} for the spin configurations analyzed 
in this paper, for which there are no poles in $\Delta_t(R)$, 
we would obtain phase disagreements on the same 
order of the ones we have found.} 
The improved spin EOB Hamiltonian~\cite{BB} will incorporate 
consistently the self-force result
(e.g., Ref.~\cite{Damour:2009sm}) and can be better constrained by new
numerical simulations.

In the spin aligned or anti-aligned case, we find that the ISCO exists
for all spin magnitudes. However, in the spin aligned case, when
$a>0.8M$, the ISCO radius (frequency) starts to increase (decrease)
with increasing $a$. This is contrary to the monotonic dependence of
the ISCO radius (frequency) on the spin magnitude in the test-particle
limit. This unusual behaviour will be overcome by the improved spin
EOB Hamiltonian of Ref.~\cite{BB}. Nevertheless, since this problem
occurs only at extreme spin magnitudes and here we have numerical
waveforms of moderate spins ($|a|<0.5M$), we choose to use this spin
EOB Hamiltonian in the current calibration.

The light ring is the unstable circular orbit of a massless particle
(such as a photon) and can be computed from the deformed EOB metric or
from $H_{\rm eff\,Kerr}(\vR,\vP,\vS_{\rm Kerr})$. As in the
nonspinning case, we do find a light ring with our spin EOB
Hamiltonian. However, in contrast to the nonspinning case, for several
values of the spin parameters (including the DD configuration) our
spin EOB Hamiltonian does not yield a maximum orbital frequency. It is
worth mentioning that if we were using only the ``Kerr" part of the
spin EOB Hamiltonian, i.e. we ignore $H_{\rm
  eff\,part}(\vR,\vP,\vsigma)$, then we do find a maximum of the
orbital frequency and its value is quite close to the light ring
position.  A more detailed study has revealed that the absence of the
maximum of the orbital frequency for the full spin EOB Hamiltonian is
due to the spin-orbit coupling term $H_{\rm
  eff\,part}(\vR,\vP,\vsigma)$ defined in Eq.~\eqref{Heff2}, which as
discussed above does not reduce to the test-particle limit prediction
at PN orders higher than 2.5PN. In the improved spin EOB
Hamiltonian~\cite{BB}, $H_{\rm eff\,part}(\vR,\vP,\vsigma)$ will be
consistent with the test-particle limit prediction at all PN orders
linear in the particle spin.  Analyses using the improved spin EOB
Hamiltonian \cite{BB} have shown a reasonable agreement between the
position of the EOB light ring and the maximum of the EOB orbital
frequency.

Quite interestingly, when the numerical and EOB waveforms are aligned at low 
frequency, as discussed in detail in Sec.~\ref{sec:NRuncertainty}, we find that the 
EOB light ring is reached at time $0.3M$ 
before the peak of the numerical $h_{22}$ amplitude. Therefore, a nice
property of the nonspinning EOB model \cite{Buonanno:2009qa,Damour2009a} holds 
also in the spinning case, i.e. the EOB light ring position is a good approximation
of the peak position of the numerical $h_{22}$ amplitude. The latter property 
will be a key ingredient in the EOB waveform model, as described later in
Sec.~\ref{sec:EOBmergerRDwaveform}.

\subsection{EOB waveform: Inspiral \& Plunge}
\label{sec:EOBinspiralwaveforms}

Having described the inspiral dynamics, we now turn to the
gravitational waveforms $h_{\ell m}$. The latter can also be employed to 
compute consistently the inspiral dynamics through 
the radiation-reaction force~\cite{Damour2009a}. 
In the nonspinning case, Refs.~\cite{Damour2009a,Buonanno:2009qa}
have shown that the resummed, factorized PN waveforms proposed 
in Ref.~\cite{DIN} are in excellent agreement with the numerical 
waveforms. In Ref.~\cite{Pan:2009} we have generalized the resummed 
factorized waveforms to include spin effects. 

The resummed waveforms are written as the product of five factors, 
\begin{equation}\label{hlm}
h_{\ell m}=h_{\ell m}^{(N,\epsilon)}\,\hat{S}_{\rm eff}^{(\epsilon)}\,T_{\ell m}\,e^{i\delta_{\ell m}}f_{\ell m}\,,
\end{equation}
where $\epsilon$ denotes the parity of the multipolar waveform. In the
circular-orbit case, $\epsilon$ is the parity of $\ell+m$:
$\epsilon=\pi(\ell+m)$.  These factors are discussed extensively in
Ref.~\cite{DIN}. Here we simply write down the expressions used in our
spin EOB model, valid for spins aligned or anti-aligned with the
orbital angular momentum. Thus, we restrict ourselves to the
equatorial plane ($\theta = \pi/2$ and $p_\theta=0$).  The leading
term $h_{\ell m}^{(N,\epsilon)}$ is the Newtonian contribution
\begin{equation}
h_{\ell m}^{(N,\epsilon)}=\frac{M\nu}{\cal{R}}n_{\ell m}^{(\epsilon)}\,c_{\ell+\epsilon}(\nu)\,v_\phi^{(\ell+\epsilon)}\,
Y^{\ell-\epsilon,-m}\,\left(\frac{\pi}{2},\Phi\right)\,,
\end{equation}
where $\cal{R}$ is the distance from the source.
The $n_{\ell m}^{(\epsilon)}$ and $c_{\ell+\epsilon}(\nu)$ are
functions given in Eqs.~(5)--(7) of Ref.~\cite{DIN}.  The $Y^{\ell
  m}(\theta,\phi)$ are the scalar spherical harmonics. The tangential
velocity $v_\phi$ is the non-Keplerian velocity of a spherical orbit
defined by $v_\phi=r_\Omega\,\hat{\Omega}$ where
\begin{equation}\label{romega}
r_\Omega\equiv\hat{\Omega}_{\rm cir}^{-2/3}=\left(M\,\left.\frac{\partial H_{\rm eff}(P_R=0)}{\partial P_\phi}\right|_{P_\phi=P_{\phi,\rm cir}}\right)^{-2/3}\,,
\end{equation}
and $P_{\phi,\rm cir}$ is the solution of the spherical orbit condition 
$\partial H_{\rm eff}(R,P_R=0,P_\phi)/\partial R=0$. As in 
the nonspinning case, the functions $\hat{S}_{\rm eff}^{(\epsilon)}$, 
$T_{\ell m}$, $e^{i\delta_{\ell m}}$ and $f_{\ell m}$ appearing in the right hand side 
of Eq.~\eqref{hlm} are computed using the Keplerian velocity
$v=\hat{\Omega}^{1/3}$. Moreover,
$\hat{S}_{\rm eff}^{(\epsilon)}$ is an effective source term that in
the test-particle, circular-motion limit contains a pole at the EOB
light ring. It is given in terms of the EOB dynamics as
\begin{equation}\label{hlmsource}
\hat{S}_{\rm eff}^{(+)}=\hat{H}_{\rm eff} \qquad \hat{S}_{\rm eff}^{(-)}=\hat{L}_{\rm eff}\equiv\left|\vrr\times\vpstar\right| \,.
\end{equation}
Setting $\hat{S}_{\rm eff}^{(-)}$ to
$\left|\vrr\times\vpstar\right|$ in Eq.~(\ref{hlmsource}) is not the
only possible choice; for example, one may instead 
choose $\hat{S}_{\rm eff}^{(-)}$ to be either $\hat{H}_{\rm eff}$ or 
$\hat{J}_{\rm eff}=\left|\vrr\times\vpstar+\vS_{\rm Kerr}/(M\mu)\right|$.
The effect of this choice
on the spin EOB model investigated in this paper is marginal, 
since in the equal-mass, equal-spin, non-precessing binary
configurations, odd parity modes contribute only a
tiny fraction of the total energy flux (see
Sec.~\ref{sec:comparinghlm} for details). Although we choose to use
the source term defined in Eq.~\eqref{hlmsource}, there is no evidence
indicating that this choice is better or worse than others for those
binary configurations in which odd parity modes are more important.

The function $T_{\ell m}$ in the right hand side of Eq.~\eqref{hlm} 
resums leading logarithms of tail effects, 
and $e^{i\delta_{\ell m}}$ is a phase correction due to subleading
logarithms. Through 2PN order, there are no tail contributions due to
spin effects and $T_{\ell m}$ and $e^{i\delta_{\ell m}}$ do not differ
from the nonspinning case. Their explicit expressions are given in
Eqs.~(19)--(29) of Ref.~\cite{DIN}. Finally, the functions $f_{\ell m}$ 
in the right hand side of Eq.~\eqref{hlm} collect the remaining PN terms.  
We computed~\cite{Pan:2009} the spin terms in $f_{\ell m}$ by Taylor expanding 
the $h_{\ell m}$ in Eq.~\eqref{hlm} and comparing it to the Taylor-expanded 
$h_{\ell m}$ calculated in PN theory, including the test-particle spin effects
through 4PN order. In the test-particle limit, we choose $\vS_{\rm
  Kerr}$ as the spin variable of the spacetime. Expressions of
$f_{\ell m}$ can be read from Ref.~\cite{Pan:2009} \footnote{For odd
  parity modes, depending on the choice of the source term among
  $\hat{H}_{\rm eff}$, $\hat{L}_{\rm eff}$ and $\hat{J}_{\rm eff}$, the
  corresponding choice of $f_{\ell m}$ should be made among the
  expressions of $f_{\ell m}^H$, $f_{\ell m}^L$ and $f_{\ell m}^J$.}.

Following Refs.~\cite{DIN,Buonanno:2009qa}, we resum all the
nonspinning terms in $f_{\ell m}$ in the functional form $ f_{\ell
  m}^{\rm NS}=(\rho_{\ell m})^\ell$ that holds at known PN orders,
where $f_{\ell m}^{\rm NS}$ collects the nonspinning terms in $f_{\ell
  m}$, and $\rho_{\ell m}$ can be read from Appendix C of
Ref.~\cite{DIN}.  The motivation for this $\rho$-resummation is to
reduce the magnitude of the 1PN coefficients in $f_{\ell m}$ that grow
linearly with $\ell$ (see Sec.~IID of Ref.~\cite{DIN}). Since at
leading order we did not find such dependence on $\ell$ in the
spinning terms~\cite{Pan:2009}, we do not apply the $\rho$-resummation
to them.

Furthermore, since we are trying to reproduce effects in the numerical
simulations that go beyond the quasi-circular motion assumption,
motivated by the PN expansion for generic orbits, we include
non-quasicircular (NQC) effects in $h_{\ell m}$ in the form
\begin{eqnarray}\label{inspwavenew}
&& {h}_{\ell m}^{\rm insp-plunge} \equiv {}^{\rm NQC}{h}_{\ell m} =
\widehat{h}_{\ell m}\,\left [ 1 + a^{h_{\ell m}}_{1}\,
\frac{p_{r^*}^{2}}{(r \,\hat{\Omega})^{2}} \right .\nonumber \\
&& \left . + a^{h_{\ell m}}_{2}\,\frac{p_{r^*}^{2}}{(r\,\hat{\Omega})^{2}}\frac{1}{r} + a^{h_{\ell m}}_{3}\,
\frac{p_{r^*}^{2}}{(r\,\hat{\Omega})^{2}}\frac{1}{r^{3/2}} \right. \nonumber \\
&& \left . + a^{h_{\ell m}}_{4}\,\frac{p_{r^*}^{2}}{(r\,\hat{\Omega})^{2}}\frac{1}{r^2} \right ]\,.
\end{eqnarray}
A similar expression was used in Ref.~\cite{Buonanno:2009qa} except that
there we used  $\dot{r}$ instead of $p_{r^*}$. For a test-particle plunging 
in the Kerr geometry, $\dot{r}$ goes to zero at the horizon. We observe 
a similar behavior in the EOB $\nu$-deformed Kerr geometry. Therefore, 
in contrast to the nonspinning case, the evolution of $\dot{r}$ is not 
monotonic during the inspiral-plunge: $\dot{r}$ increases during the inspiral, 
reaches a peak, and then starts decreasing during the plunge. 
By replacing $\dot{r}$ with $p_{r^*}$, we keep the NQC correction terms 
in Eq.~\eqref{inspwavenew} monotonic in
time; thus, they can successfully model the monotonically increasing
amplitude differences between the quasi-circular EOB and numerical waveforms.  
As in Ref.~\cite{Buonanno:2009qa}, we fix two of the four adjustable
parameters $a^{h_{22}}_{i}$ by requiring that the peaks of the
numerical and EOB $h_{22}$ waveforms coincide in both time and
amplitude, where the peak time of the numerical $h_{22}$ waveform 
is accurately predicted by the EOB light ring, as discussed above. 
The other two $a^{h_{22}}_{i}$ parameters are 
determined by minimizing the overall amplitude difference with respect
to the numerical waveform as explained in detail below.  The NQC corrections in Eq.~\eqref{inspwavenew} 
also depend on spins. However, there is not enough numerical
information in this work (we have only the DD configuration) 
to discriminate between the spinning and nonspinning 
contribution. 

\subsection{EOB waveform: Merger \& Ringdown}
\label{sec:EOBmergerRDwaveform}

The merger-ringdown waveform in the spin EOB model is built in the
same way as in the nonspinning EOB model. Details on building
merger-ringdown waveforms can be found in Sec. IIC of
Ref.~\cite{Buonanno:2009qa}. Here we briefly summarize the key points.

In the spin EOB model, the ring-down waveform is a linear combination
of eight quasinormal modes (QNMs) of the final Kerr black hole. Mass
and spin of the final black hole are computed from numerical data. In
particular, for the numerical simulation of the DD configuration, we
use $M_{\rm BH}/M=0.961109\pm0.000003$ and $a_{\rm BH}/M_{\rm
  BH}=0.54781\pm0.00001$ computed in
Ref.~\cite{Chu:2009md}. Frequencies of the QNMs are computed by
interpolating data from Ref.~\cite{Berti:2009}. The eight complex
coefficients of the linear combination are fixed by the {\it hybrid
  comb matching} described in Sec.~ IIC of Ref.~
\cite{Buonanno:2009qa}. The matching time $t_{\rm match}^{\ell m}$ is
fixed to be the EOB light-ring position. The matching interval $\Delta
t_{\rm match}^{\ell m}$ is an adjustable parameter that is fixed by
reducing the difference against numerical merger-ringdown waveforms.

\section{Calibrating the EOB waveforms to numerical relativity 
simulations}
\label{sec:EOBcalibration}

We now calibrate the spin EOB model against the numerical 
UU and DD spin configurations. We extract both the
Newman-Penrose (NP) scalars $\Psi_4^{\ell m}$ and the strain waveforms
$h_{\ell m}$ from the simulations. The strain waveforms are extracted
with the Regge-Wheeler-Zerilli (RWZ) formalism~
\cite{ReggeWheeler1957,Zerilli1970b,Sarbach2001,Rinne2008b}
(see Appendix of Ref.~\cite{Buonanno:2009qa} for details of the
numerical implementation used to obtain $h_{\ell m}$). We use
the RWZ $h_{\ell m}$ to calibrate the EOB model, and use the NP
$\Psi_4^{\ell m}$ to check the consistency between the two
wave-extraction schemes and to estimate the numerical error associated
with the wave extraction schemes.

We will use the $\ell=2, m=2$ component of the numerical waveform for
tuning the EOB model.  Thus, we calibrate in total the following six adjustable
EOB parameters: $b(\nu)$, $a^{\rm 3PN}_{\rm SS}$,
$a^{h_{22}}_{1}$, $a^{h_{22}}_{2}$, $a^{h_{22}}_{3}$ and
$\Delta t_{\rm match}^{22}$.

\subsection{Uncertainties in numerical waveforms}
\label{sec:NRuncertainty}

In this section, we compare numerical waveforms computed at different
numerical resolutions and/or using different extrapolation procedures,
or with different wave-extraction schemes. Estimates of numerical
errors in the waveforms will set our standards when calibrating the EOB
model.

First, we adopt the same waveform-alignment procedure
used in Ref.~\cite{Buonanno:2009qa}, that is we align waveforms at
low frequency by minimizing the quantity
\begin{equation}\label{waveshifts}
\Xi(\Delta t,\Delta\phi)=\int_{t_1}^{t_2}\left[\phi_1(t)-
\phi_2(t-\Delta t)-\Delta\phi\right]^2\,dt\,,
\end{equation}
over a time shift $\Delta t$ and a phase shift $\Delta\phi$, where
$\phi_1(t)$ and $\phi_2(t)$ are the phases of the two waveforms. The
range of integration $(t_1,t_2)$ is chosen to be as early as possible
to maximize the length of the waveform
but late enough to avoid the contamination from junk radiation
present in the numerical initial data. The range of integration should
also be large enough to average over numerical noise. We fix
$t_1=500M$ and $t_2=1500M$ in Eq.~\eqref{waveshifts}.
\begin{figure}
  \includegraphics[width=0.8\linewidth]{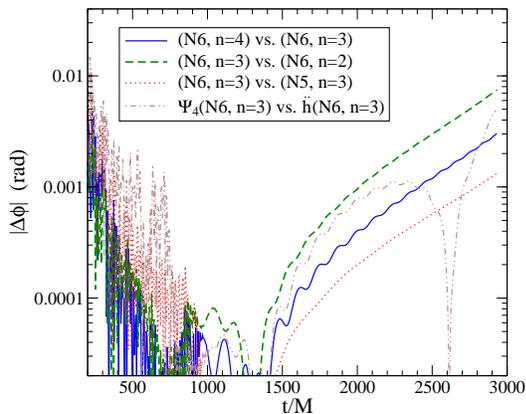}
   \caption{(color online). Numerical error estimates for the UU configuration. We show the phase difference between 
several numerical $\Psi_4^{22}$ waveforms aligned using the procedure defined by Eq.~\eqref{waveshifts}.
\label{fig:DeltaPhaseNumPsi4UU}}
\end{figure}
\begin{figure}
  \includegraphics[width=0.8\linewidth]{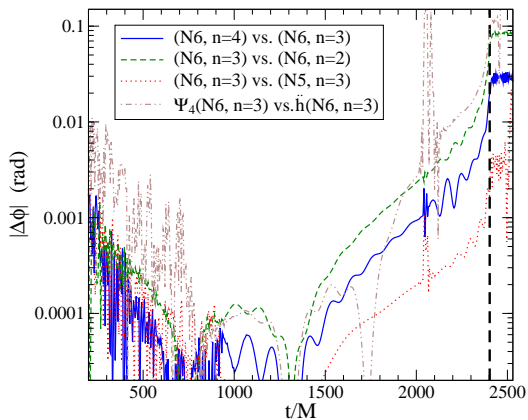}
   \caption{(color online). Numerical error estimates for the DD configuration. We show the phase difference between 
several numerical $\Psi_4^{22}$ waveforms aligned using the procedure defined by Eq.~(\ref{waveshifts}).
\label{fig:DeltaPhaseNumPsi4DD}}
\end{figure}
Using this alignment procedure, we estimate the errors on the
numerical $\Psi_4^{22}$. Figs.~\ref{fig:DeltaPhaseNumPsi4UU} and
~\ref{fig:DeltaPhaseNumPsi4DD} summarize the phase errors for
numerical $\Psi_4^{22}$. The numerical waveform labeled ``(N6, n=3)''
[or ``$\Psi_4$(N6, n=3)''] is the reference numerical waveform used
throughout this paper. Each waveform is extracted on a set of spheres
at fixed distances from the source, and then extrapolated to future
null infinity; the labels $n$ refer to different orders of this
extrapolation and are used to quantify the uncertainty in the phase
due to extrapolation. The waveform labeled by N5 (as opposed to N6) is
from a simulation with a lower numerical resolution and is used to
quantify the uncertainty due to numerical truncation errors. The
waveform labeled by ``$\ddot{h}$(N6, n=3)'' is generated by twice
differentiating the RWZ-extracted ``$h$(N6, n=3)'' waveform, and is
used to quantify the uncertainty due to the systematic difference
between extracting the NP scalar and extracting the strain waveform
via the RWZ formalism.

\begin{figure}
\includegraphics[width=\linewidth]{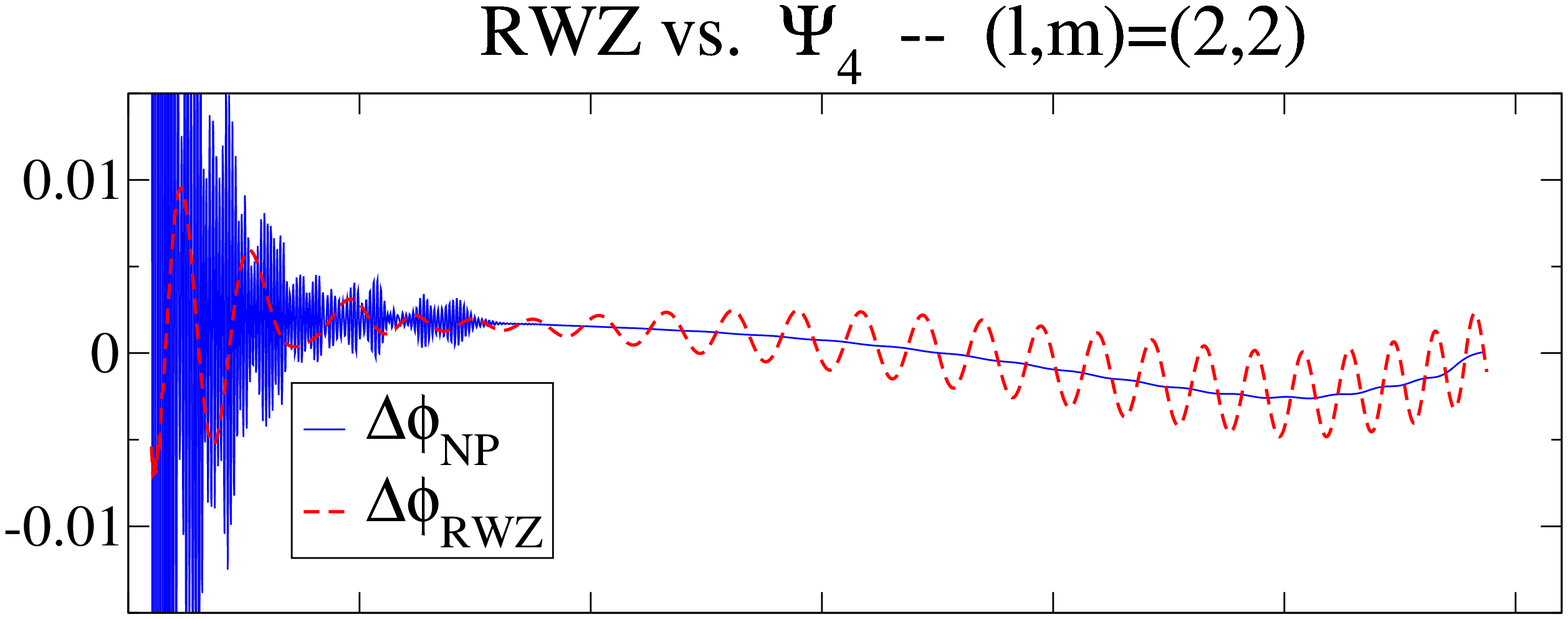}
\includegraphics[width=\linewidth]{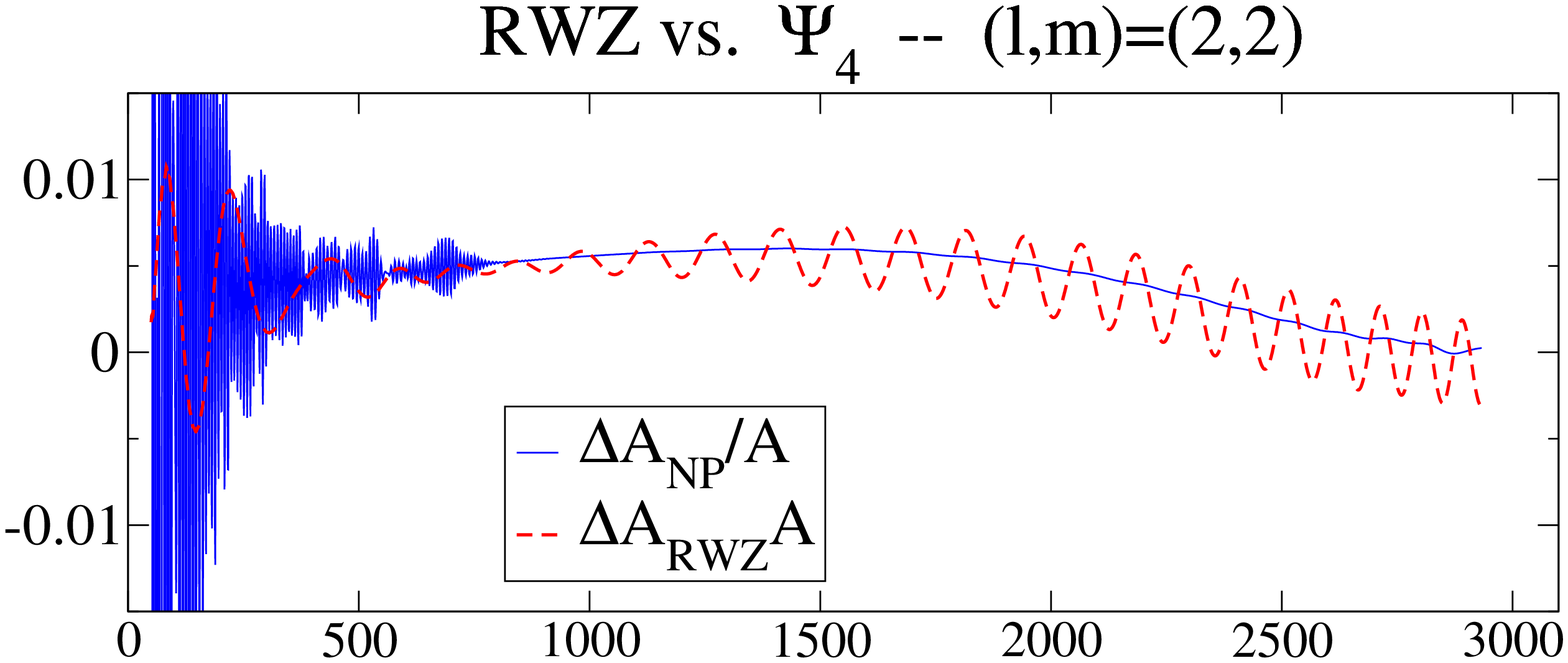}
\caption{\label{fig:RWZ-Psi4-Comparison22-UU} 
Phase and relative amplitude
  difference between the $(l,m)\!=\!(2,2)$ modes of the RWZ waveform
  $h_{\rm RWZ}$ and NP scalar $\Psi_4$ for the UU case.}
\end{figure}

\begin{figure}
\includegraphics[width=\linewidth]{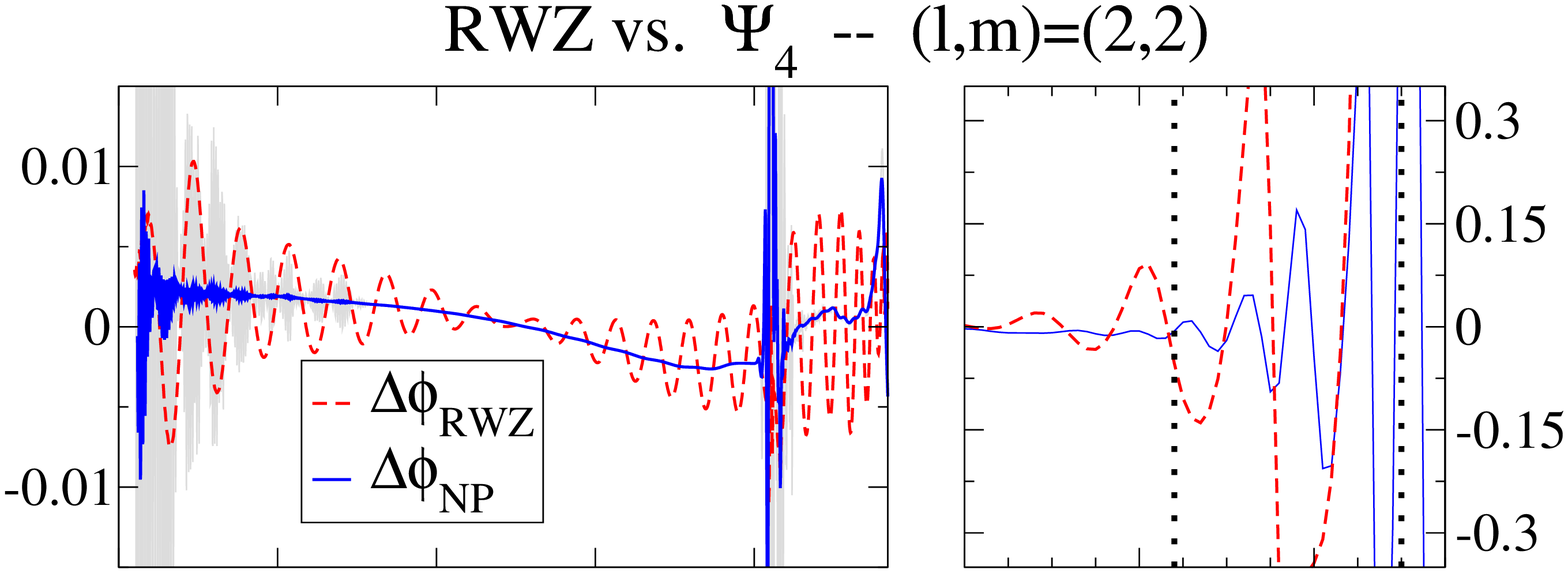}
\includegraphics[width=\linewidth]{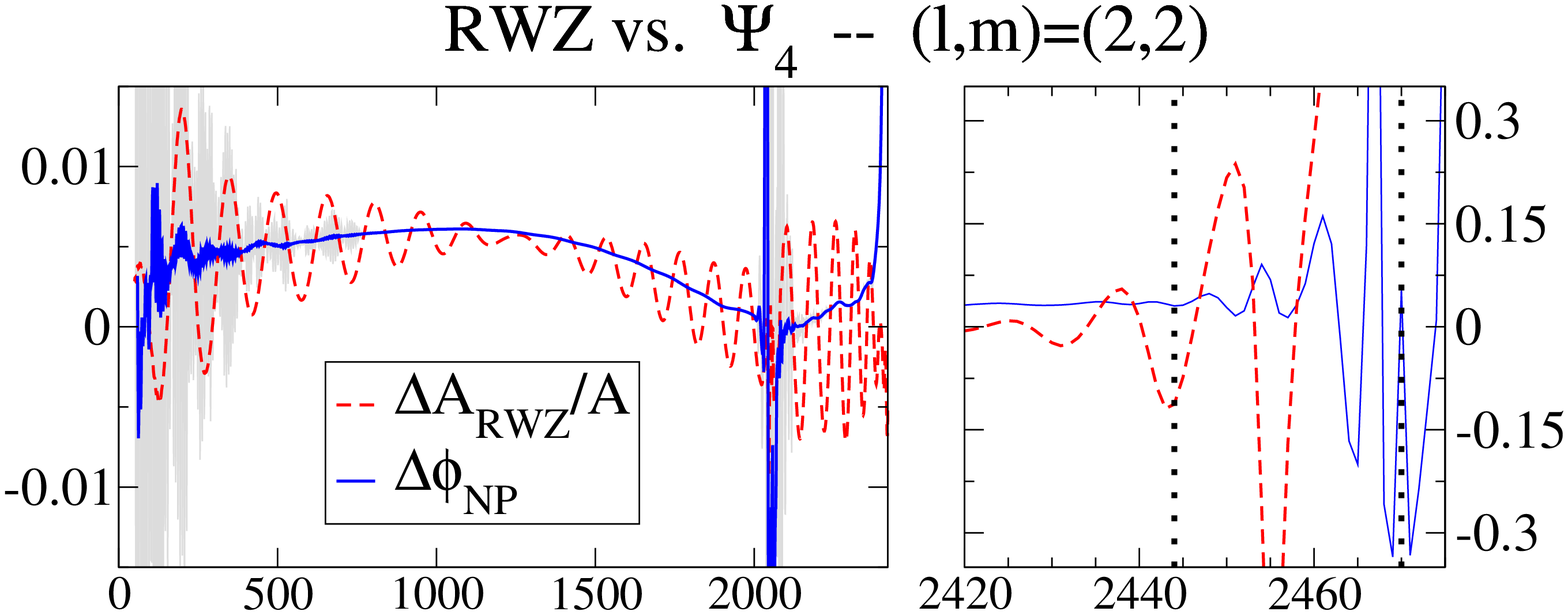}
\caption{\label{fig:RWZ-Psi4-Comparison22-DD} 
Phase and relative amplitude
  difference between the $(l,m)\!=\!(2,2)$ modes of the RWZ waveform
  $h_{\rm RWZ}$ and NP scalar $\Psi_4$ for the DD case. 
 The right
  panel shows an enlargement of merger and ringdown, with the dotted
  vertical lines indicating time of maximum of $|\Psi_4|$, and where
  $|\Psi_4|$ has decayed to 10\% and 1\% of the maximal
  value. (The blue lines are smoothed; the grey data in the background represents the unsmoothed data.)}
\end{figure}

The noise before $t=500M$ is due to spurious radiation from
initial conditions. The features around $t\approx 2100M$ in
Fig.~\ref{fig:DeltaPhaseNumPsi4DD} are due to a change of gauge in the
numerical simulation.  Extrapolation with $n=2$ leads to systematic
errors in the extrapolated waveform which in turn results in a
systematic error in $\Delta t$. Therefore, the green dashed lines in
Figs.~\ref{fig:DeltaPhaseNumPsi4UU} and \ref{fig:DeltaPhaseNumPsi4DD}
represent a possibly overly conservative error estimate. There is a
tiny frequency difference between the NP and RWZ extracted waveforms,
which is magnified into a substantial time shift when the waveforms
are aligned at low frequency. As a consequence, the dot-dashed brown
line in Fig.~\ref{fig:DeltaPhaseNumPsi4DD} shows a larger
phase difference which builds up during the late inspiral. 
It provides us with the most conservative error estimate for 
the DD configuration. This is better
illustrated in Figs.~\ref{fig:RWZ-Psi4-Comparison22-UU} and
\ref{fig:RWZ-Psi4-Comparison22-DD}, where we compare the RWZ $h_{22}$
and NP $\Psi_4^{22}$ waveforms without any time or phase shift. In
blue solid lines, we show the phase and relative amplitude differences
$\Delta\phi_{\rm NP}$ and $\Delta A_{\rm NP}/A$ between the RWZ
$h_{22}$ waveform differentiated twice with respect to time
and $\Psi_4^{22}$. In red dashed
lines, we show the phase and relative amplitude differences
$\Delta\phi_{\rm RWZ}$ and $\Delta A_{\rm RWZ}/A$ between
$\Psi_4^{22}$ integrated twice in time and the RWZ $h_{22}$. In
Fig.~\ref{fig:RWZ-Psi4-Comparison22-DD}, 
$\Delta\phi_{\rm NP}$ shows a slope between $t=500M$ and $1500M$. When
we apply the alignment procedure, this slope is removed through a time
shift, which is transformed into a larger
phase difference during late
inspiral where the wave frequency is large.


\subsection{Calibrating the EOB adjustable parameters}
\label{sec:EOBtuning}

Here we adopt the procedure suggested in Ref.~\cite{Buonanno:2009qa}, 
augmented with the iterative scheme suggested in Ref.~\cite{Damour2009a} 
when calibrating the adjustable parameters.

We divide the adjustable parameters into three groups and calibrate
them in two steps. The first group, EOB-dynamics parameters, consists
of $b(\nu)$ and $a^{\rm 3PN}_{\rm SS}$ in the EOB Hamiltonian (there
is no adjustable parameter in the model of the EOB energy flux). These
parameters determine the inspiral and plunge dynamics of the EOB model
and affect the merger-ringdown waveform only indirectly through the
waveform's phase and frequency around the matching point. The second
group, EOB-NQC parameters, consists of $a^{h_{\ell m}}_{i}$, which
enter both the EOB dynamics (through the energy flux) and the EOB
waveform (through the NQC correction). The third group, EOB-waveform
parameters, consists of $\Delta t_{\rm match}^{\ell m}$, which affect
the EOB merger-ringdown waveform but not the EOB inspiral-plunge
waveform. All the EOB adjustable parameters are calibrated to the
numerical RWZ $h_{22}$.  In the first step of calibration, we
simultaneously reduce the difference in waveforms against the
numerical UU and DD configurations by calibrating the EOB-dynamics and
the EOB-NQC adjustable parameters. In the second step, using the
adjustable parameters fixed in the first step, we calibrate the
EOB-waveform adjustable parameters.

We adopt the iterative scheme suggested in Ref.~\cite{Damour2009a} to
fix the EOB-dynamics and the EOB-NQC parameters in the first step of
calibration. In each iteration, we first minimize the phase difference
by calibrating the EOB-dynamics adjustable parameters. To be specific,
we choose to minimize the quantity $\max\limits_t\left[\phi_{\rm
    EOB}(t)-\phi_{\rm NR}(t)\right]-\min\limits_t\left[\phi_{\rm
    EOB}(t)-\phi_{\rm NR}(t)\right]$. By comparing the EOB model to
both the UU and DD configurations it is possible to calibrate the
parameters $b(\nu)$ and $a^{\rm 3PN}_{\rm SS}$ separately.  This is
because $b(\nu)$ alters a 3.5PN spin-orbit coupling term that depends
on the spin orientation, so the phases of the UU and DD waveforms
change in opposite directions when varying $b(\nu)$, but $a^{\rm
  3PN}_{\rm SS}$ alters a 3PN spin-spin coupling term, so the phases
of the UU and DD waveforms change in the same direction when varying
$a^{\rm 3PN}_{\rm SS}$.

The EOB-NQC adjustable parameters are calibrated only to the numerical
$h_{22}$ waveform of the DD configuration, because we did not run the
UU case through merger and ringdown.  We first fix $a^{h_{22}}_{1}$
and $a^{h_{22}}_{2}$ by requiring that a local extremum of the EOB
$h_{22}$ amplitude coincides with the peak of the numerical $h_{22}$
in time and amplitude (the peak time is predicted by the EOB light
ring). We expect that in the future, the peak amplitude of the
numerical $h_{22}$ will be predicted by numerical relativity with high
accuracy as an interpolation function on the physical
parameters. Therefore, $a^{h_{22}}_{1}$ and $a^{h_{22}}_{2}$ can be
determined without a least-squares fit to the NR waveform, reducing by
two the number of parameters to be determined by a least-squares
fit. The other two NQC parameters, $a^{h_{22}}_{3}$ and
$a^{h_{22}}_{4}$, are calibrated to the numerical waveform to further
reduce the disagreement in amplitude. The NQC parameters will enter
the flux through the NQC waveform ${}^{\rm NQC}{h}_{22}$ in the next
iteration. They are set to zeros initially to start the iteration and
they usually converge within five iterations.

In the third step, we calibrate the EOB-waveform adjustable 
parameter $\Delta t_{\rm match}^{22}$ by reducing the difference 
in the DD configuration merger-ringdown $h_{22}$ waveform. 

\subsection{Comparing the gravitational-wave modes \boldmath{$h_{22}$}}
\label{sec:comparingh22}

Before calibrating the EOB adjustable parameters, we investigate the 
phase difference for the EOB uncalibrated waveforms. 
For the uncalibrated model, we set $a_5=1.775$ and all six 
of our adjustable parameters to zero. We find that 
during the inspiral, the phase agreement between the numerical 
and spin EOB uncalibrated waveforms is already substantially better than 
the agreement between numerical and Taylor-expanded PN waveforms. 
For the latter, we consider the 3.5PN spin Taylor model (T4) of Ref.~
\cite{Pan2004} with amplitude corrections through
the highest PN order currently known~\cite{Kidder:2007rt,Pan:2009}. In fact, using
the uncalibrated spin EOB model and aligning the waveforms with the procedure
defined by Eq.~\eqref{waveshifts}, we find that the phase differences
against the numerical UU and DD waveforms, at the end of the simulation and at merger, 
respectively, are $-0.2$ and $4.3$ rads\footnote{If in
  the uncalibrated EOB model, we chose to include both $a_5$ and
  $a_6$, as discussed in
  Sec.~\ref{sec:EOBdynamics}, and adopted the values $a_5=-15.5$ and
  $a_6=223$ (calibrated to equal-mass nonspinning numerical waveforms
  and consistent with the constraint derived from self-force
  results in Ref.~\cite{Damour:2009sm}), we would find for the 
phase differences $-0.3$ and $3.5$ rads. They are comparable
  with the differences found in the spin EOB model with only $a_5$.}. 
Using the spin Taylor T4 model, the corresponding
phase differences are $2.0$ and $-10.0$ rads. Therefore, the spin
EOB model, even uncalibrated, improves the phase agreement with
numerical waveforms of Taylor-expanded PN models by resumming 
the PN dynamics.

\begin{figure}
  \includegraphics[width=0.9\linewidth]{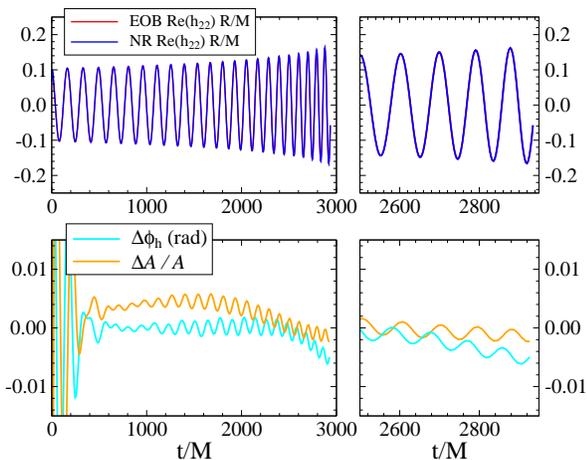}
   \caption{(color online). Comparison between the numerical and EOB waveform for the UU configuration using 
$b(\nu)=-1.65$ and $a^{\rm 3PN}_{\rm SS}=1.5$. The top panels show the real part of the numerical and EOB $h_{22}$, the bottom panels show amplitude and phase differences between them.  The left panels show times $t=0$ to $2950M$, whereas the right panels present an enlargement of the later portion of the waveform.
\label{fig:h22UUcomparisons}}
\end{figure}
\begin{figure}
  \includegraphics[width=0.9\linewidth]{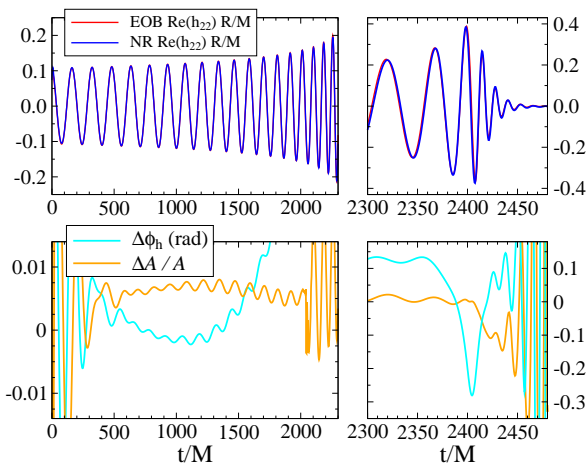}
   \caption{(color online). Comparison between the numerical and EOB waveform for the DD configuration using $b(\nu)=-1.65$ and $a^{\rm 3PN}_{\rm SS}=1.5$. The top panels show the real part of the numerical and EOB $h_{22}$, 
the bottom panels show amplitude and phase differences between them. The left panels show 
times $t=0$ to $2300M$, and the right panels show times $t=2300M$ to
$t=2480M$ on a different vertical scale.
\label{fig:h22DDcomparisons}}
\end{figure}

When calibrating the EOB model, we find that the difference in phase
and amplitude between the numerical and EOB waveforms is minimized
when fixing the EOB-dynamics parameters $b(\nu)=-1.65$ and $a^{\rm
  3PN}_{\rm SS}=1.5$.

In Fig.~\ref{fig:h22UUcomparisons}, we compare numerical and EOB
$h_{22}$ waveforms for the UU configuration. The phase difference and
relative amplitude difference are strictly within 0.01 rads and 1\%,
respectively. The systematic error in the EOB waveform in the UU
configuration is therefore smaller than the numerical errors.

In Fig.~\ref{fig:h22DDcomparisons}, we compare numerical and EOB
$h_{22}$ waveforms for the DD configuration. Using $b(\nu)=-1.65$ and
$a^{\rm 3PN}_{\rm SS}=1.5$ again, we find that the best phase and amplitude
agreement is obtained when the matching occurs at an interval of
$\Delta t_{\rm match}^{22}=3.5M$ ended at $t_{\rm
  match}^{22}=2402.0M$, which is the EOB light-ring position and is $0.3M$ 
before the merger, i.e. the peak of the numerical $h_{22}$ at $t=2402.3M$. 
The NQC parameters are $a^{h_{22}}_{1}=-16.1052$, $a^{h_{22}}_{2}=-1124.43$,
$a^{h_{22}}_{3}=4529.21$ and $a^{h_{22}}_{4}=-4587.53$. The relative
amplitude difference is strictly within $1\%$ until $2000M$. After
$2000M$, although oscillations due to numerical gauge effects in the RWZ
$h_{22}$ waveform are at the level of $2\%$ until the merger, the average
difference is still less than $1\%$. After the merger, the amplitude
difference grows to about $-5\%$ and starts oscillating with increasing
magnitude. The latter phenomenon is due to gauge effects in the RWZ
$h_{22}$ waveform as discussed in Sec.~\ref{sec:NRuncertainty} and the
Appendix of Ref.~\cite{Buonanno:2009qa}. The phase difference is
within 0.01 rads until about $1800M$ and grows to $-0.28$ rads
until merger and settles to about $0.1$ rads before the exponentially
decaying amplitude results in increased errors in the extracted gravitational-wave 
phase. 

In the spin DD configuration, the error in the EOB waveform
(especially in the phase) is not within the numerical errors. The
phase difference of $\sim 0.13$ rads at late inspiral around $t=2300M$
can be reduced to within the numerical errors of $\sim 0.01$ rads by
calibrating the EOB-dynamics adjustable parameters, i.e. $b(\nu)$ and
$a^{\rm 3PN}_{\rm SS}$. However, this leads to an increase of the
phase difference around the merger. Since we choose to minimize the
span of the phase difference over the range of the full inspiral,
merger and ringdown waveform, the phase difference at late inspiral is
larger than what it could have been. The largest phase difference
around merger can not be removed by calibrating the chosen adjustable
parameters. Nevertheless, we can substantially reduce the phase
difference if we allow one of the EOB-dynamics parameters $b(\nu)$ and
$a^{\rm 3PN}_{\rm SS}$ to be different in the UU and DD cases, or if
we add one more spin-independent adjustable parameter. For instance,
there can be a NQC correction factor on the right-hand side of
Eq.~\eqref{inspwavenew} that contributes solely to the phase of the
waveform in the form of~\cite{Damour2007}
\begin{equation}
h_{\ell m}^{\rm insp-plunge}={}^{\rm NQC}{h}_{\ell m}\,e^{i\,b_1^{h_{\ell m}}\,p_{r^*}/(r\hat{\Omega})}.
\end{equation}
We can reduce the phase difference at merger to $<0.05$ rads by
calibrating this extra EOB-NQC adjustable parameter $b_1^{h_{\ell
    m}}$. However, until we study a larger number of waveforms, we
will not over-tune parameters, since the main purpose of this
preliminary and exploratory study on the spin EOB model is to show
that by making a very simple and minimal choice of adjustable
parameters, we can achieve a quite fair agreement with the numerical
simulations.

We shall emphasize that, despite the small phase difference that
exceeds the numerical errors in the DD configuration, the faithfulness
of the EOB waveforms with the numerical waveforms is very good. Using
the noise curves of Enhanced LIGO and Advanced LIGO\footnote{ For
  Enhanced LIGO, we use the power spectral density given at
  \url{http://www.ligo.caltech.edu/~rana/NoiseData/S6/DCnoise.txt};
  for Advanced LIGO, we use the broadband configuration power spectral
  density given at
  \url{http://www.ligo.caltech.edu/advLIGO/scripts/ref_des.shtml}.},
for both the UU and DD configurations, we find that the faithfulness
is always better than 0.999 for black-hole binaries with a total mass
of $30\mbox{--}200M_\odot$. Note that the numerical waveforms
  start roughly at $40$Hz for binaries with total mass $30M_\odot$ and
  at $10$Hz for binaries with total mass $100M_\odot$. Since the
  Advanced LIGO noise curve has a low frequency cutoff at $10$Hz, the
  numerical waveforms are not long enough to cover the Advanced LIGO
  sensitivity band for binaries with total mass smaller than
  $100M_\odot$. When computing overlaps for these lower mass binaries
  using Advanced LIGO noise curve, we start the integration at the
  starting frequency of the numerical waveforms instead of $10$ Hz. 
  For the available numerical waveforms, the overlaps with EOB
  waveforms are well above the
requirement on the accuracy of binary black-hole waveforms
for detection and measurement purposes in gravitational-wave
observations~\cite{Lindblom2008}.

\begin{figure}
  \includegraphics[width=0.8\linewidth]{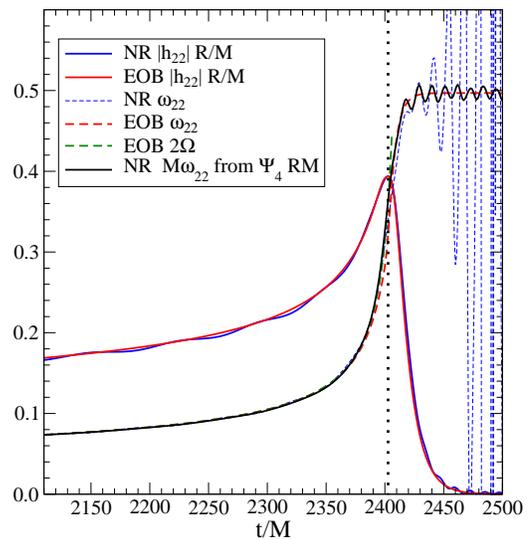}
  \caption{(color online) We show the amplitude and frequency of the
    numerical and EOB mode $h_{22}$, the EOB orbital frequency and the
    frequency of the numerical mode $\Psi_4^{22}$ for the DD
      configuration. The vertical line marks the peak of the
    amplitude of the numerical waveform. The EOB light-ring is $0.3M$
    before the peak and is too close to be shown in the figure.
\label{fig:h22DDAmpOmega}}
\end{figure}
In Fig.~\ref{fig:h22DDAmpOmega}, we compare the amplitude and
frequency of numerical and EOB $h_{22}$ waveforms together with the
orbital frequency of the EOB model, for the DD configuration. 
Unlike the nonspinning case
\cite{Buonanno:2009qa}, the orbital frequency $\Omega$ continues to grow during 
the plunge. However, the EOB light-ring is very close to the peak of the numerical $h_{22}$, 
as discussed in Sec.~\ref{sec:EOBdynamics}.
Note that during the ringdown, the frequency computed from the numerical
 $h_{22}$ shows increasingly large oscillations. We also plot the frequency 
computed from the numerical $\Psi_4^{22}$ model. This frequency shows much smaller, and
bounded, oscillations deep into the ringdown regime.

\subsection{Comparing the gravitational-wave modes \boldmath{$h_{\ell m}$}}
\label{sec:comparinghlm}

Here we generate inspiral higher-order modes, $h_{\ell m}$, using the same 
dynamics-adjustable parameters calibrated to the numerical $h_{22}$ mode in the previous
section. The EOB-NQC parameters and the EOB-waveform parameters for
these modes are not calibrated, since higher-order numerical waveforms
show large numerical errors before reaching their peaks. For this reason, 
we constrain the comparison between numerical and EOB
higher-order modes to the inspiral stage. The higher-order modes are aligned at low 
frequencies using the same time and phase shifts (modulo a factor of $m/2$ in the
phase shifts) applied to the EOB $h_{22}$ mode. 

\begin{figure*}
  \includegraphics[width=0.45\linewidth]{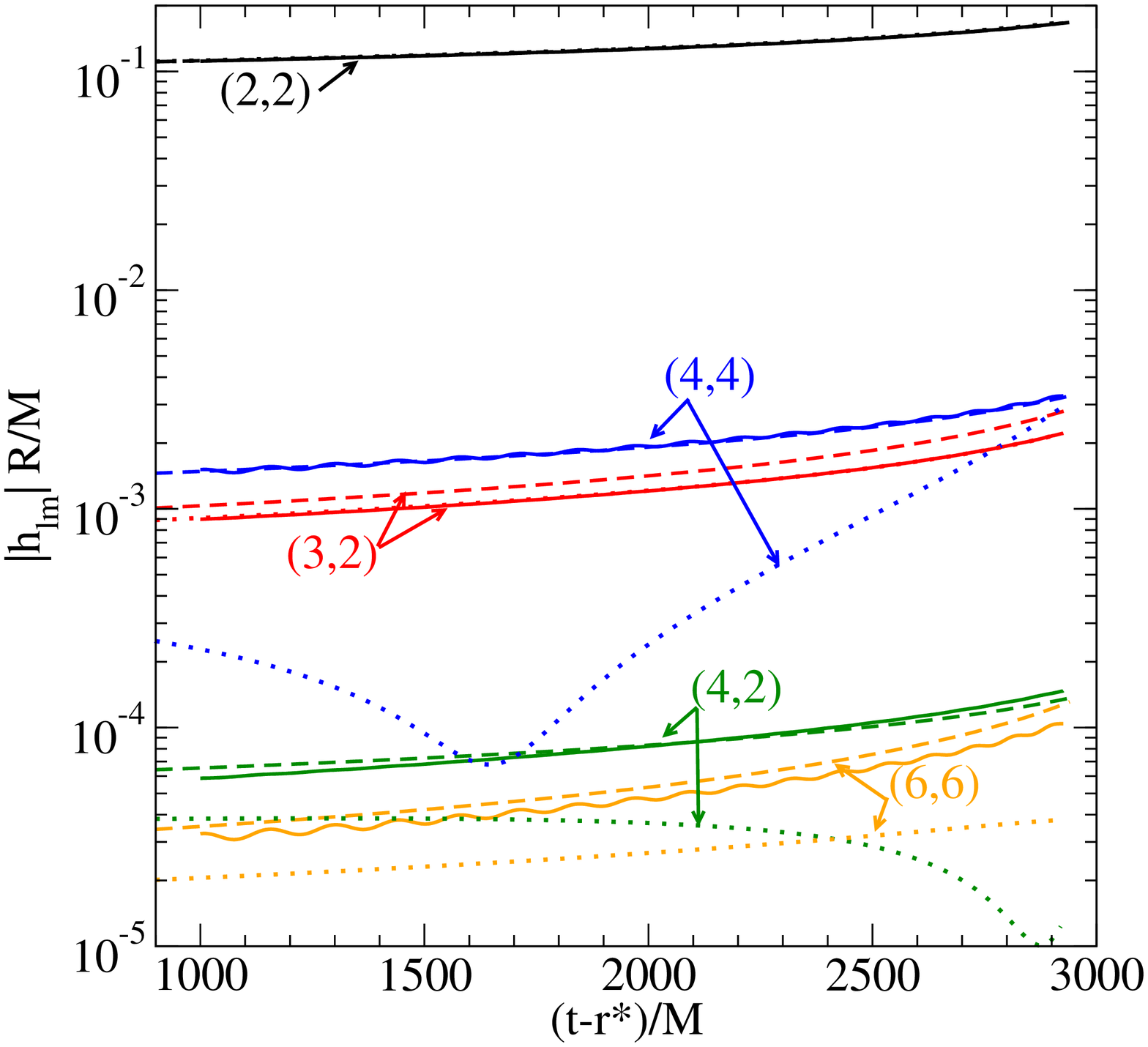} 
  \hspace{1.0truecm}
  \includegraphics[width=0.45\linewidth]{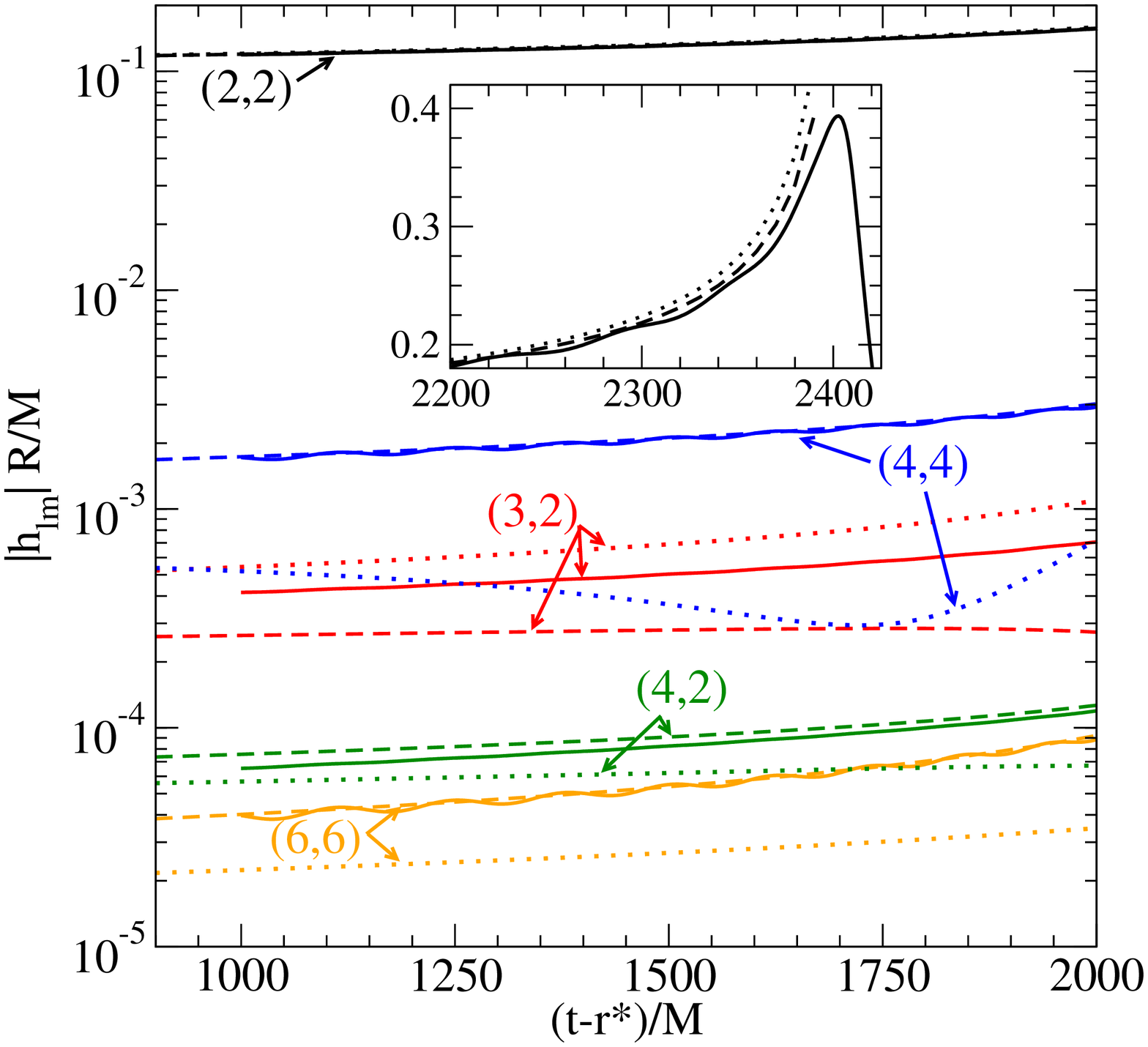}
  \caption{\label{fig:amplitudeSh_lm} (color online).  Comparison of
    the numerical (solid lines), EOB (dashed lines) and
    Taylor-expanded (dotted lines) amplitudes of the dominant and
    leading subdominant $(l,m)$ modes for the UU (left panel) and DD
    (right panel) configurations.  The inset shows the amplitudes for
    the dominant $(2,2)$ mode during the late-inspiral and plunge in
    the DD configuration, without the addition of EOB-NQC and
    EOB-waveform adjustable parameters.}
\end{figure*}

In Fig.~\ref{fig:amplitudeSh_lm}, we compare the EOB (dashed lines)
and numerical (solid lines) amplitudes of the first five modes that
dominate the signal power. In the DD case we show results only 
until $t = 2000 M$ because at later times the numerical data are 
affected by large oscillations, likely due to gauge effects. 
Except for the $h_{32}$ mode in the DD
  configuration, the agreement is very good for all the subdominant
  modes, as well as for the $h_{22}$ mode, in both the UU and DD
  waveforms. We believe that the difference seen for the DD $h_{32}$
  mode is due to the lack of knowledge of PN spin couplings in the
amplitude of the higher modes.  In fact, only the leading-order PN
spin term is known in the amplitude of the $h_{32}$ mode, and no PN
spin terms are known in the amplitudes of the other subdominant modes
shown in Fig.~\ref{fig:amplitudeSh_lm}. Being resummed in
the form of Eq.~\eqref{hlm}, the leading spin term in $h_{32}$ leaves
a residual term in $f_{32}$ at the leading order.  We test two choices
of the odd-parity source term in Eq.~\eqref{hlmsource}, $\hat{L}_{\rm
  eff}$ and $\hat{H}_{\rm eff}$, and find this residual term always
dominating over all the nonspinning terms and causing $f_{32}$ to
decrease and cross zero at high frequency, thus showing the odd
behavior in the DD (3,2) mode of Fig.~\ref{fig:amplitudeSh_lm}. We
also try to apply the $\rho$-resummation discussed in
Sec.~\ref{sec:EOBinspiralwaveforms} on the spin terms of the
$f_{32}$. Although when applying the $\rho$-resummation, the leading
order residual term in $\rho_{32}$ is reduced by $1/\ell=1/3$ with
respect to the residual term in $f_{32}$, it still dominates over other
terms and causes $(\rho_{32})^3$ to cross zero at high
frequency.

In Fig.~\ref{fig:amplitudeSh_lm}, we also show the Taylor-expanded PN
amplitudes (dotted lines). Their expressions can be read from 
Refs.~\cite{Kidder:2007rt} and \cite{Pan:2009}, and they depend on 
dynamical variables only through the orbital velocity. We calculate these 
amplitudes using the non-Keplerian orbital velocity defined in 
Eq.~\eqref{romega} for the leading term and the Keplerian 
orbital velocity for all the next-to-leading terms. We calculate 
the non-Keplerian and Keplerian velocities using the EOB dynamics. 
That is to say, these amplitudes and the resummed amplitudes are calculated 
using exactly the same dynamical evolutions. In particular, the energy 
flux in the dynamics is always modeled by resummed waveforms, even when 
we calculate the Taylor-expanded PN amplitudes.
These Taylor-expanded PN amplitudes are not to be confused with the 
amplitudes of the adiabatic PN approximants, such as the TaylorT1 and TaylorT4 
approximants \cite{Boyle2007}, because the underlying dynamics of the latter 
is completely different. In Fig.~\ref{fig:amplitudeSh_lm}, although the Taylor-expanded
PN amplitudes work reasonably well for the $h_{22}$ mode during inspiral, and probably by chance
also for the $h_{32}$ mode in the UU configuration, their performance
is not as good as that of the resummed amplitudes in 
general. Especially, for the $h_{44}$ and $h_{42}$ modes, the Taylor
amplitudes are not monotonic. This unpleasant behavior is caused by
their 1PN order non-spinning terms. Furthermore, the insert of Fig.~\ref{fig:amplitudeSh_lm} 
shows that the performance of Taylor-expanded PN amplitudes 
becomes worse for the $h_{22}$ mode during the late inspiral and 
plunge in the DD configuration.

Given the current information from PN theory and numerical simulations, 
we consider the agreement in Fig.~\ref{fig:amplitudeSh_lm} reasonable 
and do not dwell further on the choice of the waveform modeling options. 
The differences have little impact on the EOB model since the largest 
difference in $h_{32}$ affects the energy flux by less than $10^{-4}$, 
which is overwhelmed by other uncertainties in the EOB dynamics.

For the five dominant modes, the relative differences between the numerical 
and EOB $h_{\ell m}$ frequencies are within $0.5\%$, except for the $(3,2)$ mode 
in the DD configuration where the difference is within $1\%$. 
Since the $h_{\ell m}$ frequency depends on both the orbital frequency 
and its amplitude, the larger amplitude difference in the 
$(3,2)$ mode affects its gravitational-wave frequency.
Except for the $(3,2)$ mode in the DD configuration, all frequency agreement is 
within the numerical errors.

\section{Conclusions}
\label{sec:conclusions}

In this paper, we carried out the first calibration of the spin EOB 
model to accurate numerical-relativity simulations of spinning, non-precessing 
black-hole binaries. We focused on two equal-mass black-hole binaries 
having spins both aligned, or both anti-aligned with the orbital 
angular momentum, and dimensionless magnitude $\sim 0.44$~\cite{Chu:2009md}.

For the EOB conservative dynamics, we adopted the spin EOB Hamiltonian 
suggested in Refs.~\cite{Damour01c,DJS08}, augmented with the 4PN order 
non-spinning parameter $a_5$ and two adjustable parameters. For the EOB non-conservative dynamics,  
we employed the gravitational-wave energy flux which includes spin effects and 
which has been computed using the factorized multipolar waveforms of Ref.~\cite{Pan:2009}. 

As in previous cases~\cite{Boyle2008a,Buonanno:2009qa}, we aligned the
EOB and numerical waveforms at low frequency over a time interval of
$1000 M$, and minimized the difference between numerical and EOB
waveforms by calibrating a handful of EOB-adjustable parameters.  In
particular, in this first exploration, we calibrated two EOB-dynamics
adjustable parameters [$b(\nu)$ in Eq.~\eqref{bnu} which introduces a
spin-orbit term at 3.5PN order, and $a^{\rm 3PN}_{\rm SS}$ in
Eq.~\eqref{HeffSS3PN} which introduces a 3PN spin-spin term], and three
EOB-NQC adjustable parameters [see Eq.~\eqref{inspwavenew}] which
enter the gravitational-wave energy flux and the EOB
gravitational-wave (2,2) mode. Finally, we also calibrated the
EOB-waveform adjustable parameter $\Delta t_{\rm match}^{22 }$.  Quite
interestingly, similar to the case of non-spinning waveforms, we found
that for spinning waveforms, once the EOB-dynamics adjustable
parameters are calibrated at low frequency, the EOB light-ring
coincides with the peak of the numerical-relativity waveform. Thus,
for both spinning and non-spinning binary black holes, the EOB
light-ring marks the most natural point at which to match the EOB
inspiral-plunge waveform to the EOB merger-ringdown waveform.

In the equal-mass, spin aligned case, we found that phase and fractional 
amplitude differences between the numerical and EOB $(2,2)$ 
mode can be reduced to $0.01$ rads and $1\%$, respectively,
over the entire inspiral waveforms. In the equal-mass, spin anti-aligned case,
these differences can be reduced to $0.13$ rads and $1\%$ during
inspiral, and to $0.4$ rads and $10\%$ during merger and ringdown. The
waveform agreement is within numerical errors in the spin aligned
case while slightly over numerical errors in the spin anti-aligned case. 
Despite this difference, we found that using Enhanced LIGO and Advanced LIGO 
noise curves, the overlap maximized with respect to reference time and phase 
between the EOB and the numerical (2,2) mode, is larger than
0.999 for binaries with total mass $30\mbox{--}200 M_\odot$. 
This is well above the accuracy requirement of binary black-hole 
waveforms for detection and measurement purposes in gravitational-wave
observations~\cite{Lindblom2008}.

In addition to comparing the numerical and EOB waveforms for the
leading (2,2) mode, we also compared them for the next four subleading
modes. Except for the $h_{32}$ mode in the DD configuration, the amplitude 
and frequency agreements are very good for all the subdominant
  modes, as well as for the $h_{22}$ mode, in both the UU and DD
  waveforms. We believe that the difference seen for the DD $h_{32}$
  mode is due to the lack of knowledge of PN spin couplings in the
amplitude of the subleading modes. 

The spin EOB Hamiltonian~\cite{Damour01c,DJS08} adopted in this paper
was an excellent starting point to explore the calibration of the EOB
model against spinning numerical simulations; however, as discussed
above, and in particular in Sec.~\ref{sec:EOBstudydynamics}, the spin
EOB Hamiltonian we used exhibits some unusual behaviour.
Especially when extended 
at 4PN and 5PN, in some regions of
the parameter space the Hamiltonian does not have an ISCO or the ISCO
radius grows as the spin magnitude increases. This is opposite to the
result in the test-particle limit case. Moreover, although the spin
EOB Hamiltonian has a light ring, in some regions of the parameter
space (including the anti-aligned case discussed in this paper) the
orbital frequency does not reach a maximum. Those features turned out
to be crucial when calibrating the EOB model to nonspinning numerical
waveforms, and we believe will be crucial also when modeling spinning
numerical waveforms. We found that the lack of those features in the
current EOB Hamiltonian is due to the {\em ad hoc} spin coupling term
$H_{\rm eff\,part}(\vR,\vP,\vsigma)$, defined in
Eq.~\eqref{Heff2}. This spin coupling term does not reproduce the
results of a spinning test-particle at PN orders higher than 2.5PN.
Analyses using an improved spin EOB Hamiltonian~\cite{BB}
obtained by building on Ref.~\cite{BRB} have shown that those 
features can be recovered.
 
\begin{acknowledgments}
  We thank Enrico Barausse for several useful discussions, and
  Emanuele Berti for providing us with the quasi-normal mode
  frequencies and decay times used in this paper. We thank Fan Zhang
  for extrapolating the numerical waveforms to infinity. A.B. and
  Y.P. acknowledge support from NSF Grants No. PHYS-0603762 and
  PHY-0903631.  A.B. also acknowledges support from NASA grant
  NNX09AI81G.  L.B., T.C., L.K., H.P., and M.S. are supported in part
  by grants from the Sherman Fairchild Foundation to Caltech and
  Cornell, and from the Brinson Foundation to Caltech; by NSF Grants
  No. PHY-0601459 and No. PHY-0652995 at Caltech; by NASA Grant
  NNX09AF97G at Caltech; by NSF Grants No. PHY-0652952 and
  No. PHY-0652929 at Cornell; and by NASA Grant No. NNX09AF96G at
  Cornell.  H.P. gratefully acknowledges support from the NSERC of
  Canada, from Canada Research Chairs Program, and from the Canadian
  Institute for Advanced Research.
\end{acknowledgments}

\appendix

\section{Tortoise coordinate in Cartesian implementation}
\label{sec:tortoise}

We start with the definition of the radial tortoise coordinate given
in Eq.~\eqref{tortoise}: $dR^*/dR=1/\xi_a(R)$. The invariance of the
action gives $P_{R^*}=P_R\,dR/dR^*=P_R\,\xi_a(R)$.  In evolving the
EOB dynamics, we adopt the dynamical variables $\vR$, $\vPstar$,
$\vSa$ and $\vSb$.  The transform from $\vP$ to $\vPstar$ is a
coordinate transform, not a canonical transform.  In this section, we
derive explicitly the transform to tortoise coordinate for the
Hamiltonian and Hamilton equations of motion implemented in Cartesian
coordinates.

The transform between $\vP$ and $\vPstar$ is determined by the
invariance in their tangential components and the rescaling in their
radial components, that is
\begin{eqnarray}
\vR\times\vP&=&\vR\times\vPstar\,, \nonumber\\
\xi_a(R)\,\vR\cdot\vP&=&\vR\cdot\vPstar\,.
\end{eqnarray}
Choosing three independent equations out of the four above, we can write the transform in components as
\begin{eqnarray}
&&\left(\begin{array}{ccc} -Y & X & 0 \\ 0 & -Z & Y \\ X & Y & Z \end{array}\right)
\left(\begin{array}{c} P^*_X \\ P^*_Y \\ P^*_Z \end{array}\right) = \nonumber \\
&& \left(\begin{array}{ccc} -Y & X & 0 \\ 0 & -Z & Y \\ \xi_a(R)X & \xi_a(R)Y & \xi_a(R)Z \end{array}\right)
\left(\begin{array}{c} P_X \\ P_Y \\ P_Z \end{array}\right)\,,
\end{eqnarray}
or explicitly as
\begin{widetext}
\begin{equation}
\vPstar=\left(\begin{array}{c} P^*_X \\ P^*_Y \\ P^*_Z \end{array}\right) 
= \left(\begin{array}{ccc} 
1+\frac{X^2}{R^2}\,\left[\xi_a(R)-1\right] & \frac{XY}{R^2}\,\left[\xi_a(R)-1\right] & \frac{XZ}{R^2}\,\left[\xi_a(R)-1\right] \\ 
\frac{XY}{R^2}\,\left[\xi_a(R)-1\right] & 1+\frac{Y^2}{R^2}\,\left[\xi_a(R)-1\right] & \frac{YZ}{R^2}\,
\left[\xi_a(R)-1\right] \\ 
\frac{XZ}{R^2}\,\left[\xi_a(R)-1\right] & \frac{YZ}{R^2}\,\left[\xi_a(R)-1\right] & 1+\frac{Z^2}{R^2}\,\left[\xi_a(R)-1\right] 
\end{array}\right)
\left(\begin{array}{c} P_X \\ P_Y \\ P_Z \end{array}\right)\equiv T\vP\,.
\end{equation}
\end{widetext}
In the spin EOB Hamiltonian, we shall replace $\vP$ with $T^{-1}\vPstar$. The equations of motion for $\vR$ and $\vPstar$ are
\begin{equation}
\label{eomprstar1}
\frac{dX^i}{dt}
=\left.\frac{\partial H_{\rm real}}{\partial P_i}\right|_{X^i}
=\left.\frac{\partial H_{\rm real}}{\partial P^*_j}\right|_{X^j}\,\frac{\partial P^*_j}{\partial P_i}
=\left.\frac{\partial H_{\rm real}}{\partial P^*_j}\right|_{X^j}\,{T_j}^i\,,
\end{equation}
and
\begin{eqnarray}
\label{eomprstar2}
\frac{dP^*_i}{dt}
&=&\frac{\partial P^*_i}{\partial P_j}\,\frac{dP_j}{dt}+\frac{\partial P^*_i}{\partial X^j}\,\frac{dX^j}{dt}
\nonumber\\
&=&{T_i}^j\,\left(-\left.\frac{\partial H_{\rm real}}{\partial X^j}\right|_{P_j}+\frac{1}{\Omega|\vL|}\,\frac{dE}{dt}\,P_j\right)
\nonumber \\ && +\frac{\partial P^*_i}{\partial X^j}\left.\frac{\partial H_{\rm real}}{\partial P^*_k}\right|_{X^k} {T_k}^j \nonumber\\
&=&-{T_i}^j\,\left.\frac{\partial H_{\rm real}}{\partial X^j}\right|_{P_j}+\frac{1}{\Omega|\vL|}\,\frac{dE}{dt}\,P^*_i
\nonumber \\ && +\frac{\partial P^*_i}{\partial X^j}\left.\frac{\partial H_{\rm real}}{\partial P^*_k}\right|_{X^k}\,{T_k}^j \,, \
\end{eqnarray}
where the matrix $\partial P^*_i/\partial X^j$ can be written in $T$ and $\vPstar$ as
${\partial P^*_i}/{\partial X^j}={\partial {T_i}^k}/{\partial X^j}\,{\left(T^{-1}\right)_k}^l\,P^*_l$.
%\bibliography{References/References}

\begin{thebibliography}{41}
\expandafter\ifx\csname natexlab\endcsname\relax\def\natexlab#1{#1}\fi
\expandafter\ifx\csname bibnamefont\endcsname\relax
  \def\bibnamefont#1{#1}\fi
\expandafter\ifx\csname bibfnamefont\endcsname\relax
  \def\bibfnamefont#1{#1}\fi
\expandafter\ifx\csname citenamefont\endcsname\relax
  \def\citenamefont#1{#1}\fi
\expandafter\ifx\csname url\endcsname\relax
  \def\url#1{\texttt{#1}}\fi
\expandafter\ifx\csname urlprefix\endcsname\relax\def\urlprefix{URL }\fi
\providecommand{\bibinfo}[2]{#2}
\providecommand{\eprint}[2][]{\url{#2}}

\bibitem[{\citenamefont{Barish and Weiss}(1999)}]{Barish:1999}
\bibinfo{author}{\bibfnamefont{B.~C.} \bibnamefont{Barish}} \bibnamefont{and}
  \bibinfo{author}{\bibfnamefont{R.}~\bibnamefont{Weiss}},
  \bibinfo{journal}{Phys. Today} \textbf{\bibinfo{volume}{52}},
  \bibinfo{pages}{44} (\bibinfo{year}{1999}).

\bibitem[{\citenamefont{Waldman}(2006)}]{Waldman:2006}
\bibinfo{author}{\bibfnamefont{S.~J.} \bibnamefont{Waldman}}
  (\bibinfo{collaboration}{LIGO Scientific Collaboration}),
  \bibinfo{journal}{Class.\ Quantum Grav.} \textbf{\bibinfo{volume}{23}},
  \bibinfo{pages}{S653} (\bibinfo{year}{2006}).

\bibitem[{\citenamefont{Acernese et~al.}(2006)}]{Acernese-etal:2006}
\bibinfo{author}{\bibfnamefont{F.}~\bibnamefont{Acernese}} \bibnamefont{et~al.}
  (\bibinfo{collaboration}{Virgo Collaboration}), \bibinfo{journal}{Class.\
  Quantum Grav.} \textbf{\bibinfo{volume}{23}}, \bibinfo{pages}{S635}
  (\bibinfo{year}{2006}).

\bibitem[{\citenamefont{Schutz}(2009)}]{schutz_lisa_science}
\bibinfo{author}{\bibfnamefont{B.~F.} \bibnamefont{Schutz}},
  \bibinfo{journal}{Class.\ Quantum Grav.} \textbf{\bibinfo{volume}{26}},
  \bibinfo{pages}{094020} (\bibinfo{year}{2009}).

\bibitem[{\citenamefont{Buonanno
  et~al.}(2007{\natexlab{a}})\citenamefont{Buonanno, Cook, and
  Pretorius}}]{Buonanno-Cook-Pretorius:2007}
\bibinfo{author}{\bibfnamefont{A.}~\bibnamefont{Buonanno}},
  \bibinfo{author}{\bibfnamefont{G.~B.} \bibnamefont{Cook}}, \bibnamefont{and}
  \bibinfo{author}{\bibfnamefont{F.}~\bibnamefont{Pretorius}},
  \bibinfo{journal}{Phys.\ Rev.\ D} \textbf{\bibinfo{volume}{75}},
  \bibinfo{eid}{124018} (\bibinfo{year}{2007}{\natexlab{a}}).

\bibitem[{\citenamefont{Pan et~al.}(2008)\citenamefont{Pan, Buonanno, Baker,
  Centrella, Kelly, McWilliams, Pretorius, and van Meter}}]{Pan2007}
\bibinfo{author}{\bibfnamefont{Y.}~\bibnamefont{Pan}},
  \bibinfo{author}{\bibfnamefont{A.}~\bibnamefont{Buonanno}},
  \bibinfo{author}{\bibfnamefont{J.~G.} \bibnamefont{Baker}},
  \bibinfo{author}{\bibfnamefont{J.}~\bibnamefont{Centrella}},
  \bibinfo{author}{\bibfnamefont{B.~J.} \bibnamefont{Kelly}},
  \bibinfo{author}{\bibfnamefont{S.~T.} \bibnamefont{McWilliams}},
  \bibinfo{author}{\bibfnamefont{F.}~\bibnamefont{Pretorius}},
  \bibnamefont{and} \bibinfo{author}{\bibfnamefont{J.~R.} \bibnamefont{van
  Meter}}, \bibinfo{journal}{Phys.\ Rev.\ D} \textbf{\bibinfo{volume}{77}},
  \bibinfo{eid}{024014} (\bibinfo{year}{2008}).

\bibitem[{\citenamefont{Ajith et~al.}(2007)\citenamefont{Ajith, Babak, Chen,
  Hewitson, Krishnan, Whelan, Br{\"u}gmann, Diener, Gonzalez, Hannam
  et~al.}}]{Ajith-Babak-Chen-etal:2007}
\bibinfo{author}{\bibfnamefont{P.}~\bibnamefont{Ajith}},
  \bibinfo{author}{\bibfnamefont{S.}~\bibnamefont{Babak}},
  \bibinfo{author}{\bibfnamefont{Y.}~\bibnamefont{Chen}},
  \bibinfo{author}{\bibfnamefont{M.}~\bibnamefont{Hewitson}},
  \bibinfo{author}{\bibfnamefont{B.}~\bibnamefont{Krishnan}},
  \bibinfo{author}{\bibfnamefont{J.~T.} \bibnamefont{Whelan}},
  \bibinfo{author}{\bibfnamefont{B.}~\bibnamefont{Br{\"u}gmann}},
  \bibinfo{author}{\bibfnamefont{P.}~\bibnamefont{Diener}},
  \bibinfo{author}{\bibfnamefont{J.}~\bibnamefont{Gonzalez}},
  \bibinfo{author}{\bibfnamefont{M.}~\bibnamefont{Hannam}},
  \bibnamefont{et~al.}, \bibinfo{journal}{Class.\ Quantum Grav.}
  \textbf{\bibinfo{volume}{24}}, \bibinfo{pages}{S689} (\bibinfo{year}{2007}).

\bibitem[{\citenamefont{Ajith et~al.}(2008)\citenamefont{Ajith, Babak, Chen,
  Hewitson, Krishnan, Sintes, Whelan, Br\"{u}gmann, Diener, Dorband
  et~al.}}]{Ajith-Babak-Chen-etal:2007b}
\bibinfo{author}{\bibfnamefont{P.}~\bibnamefont{Ajith}},
  \bibinfo{author}{\bibfnamefont{S.}~\bibnamefont{Babak}},
  \bibinfo{author}{\bibfnamefont{Y.}~\bibnamefont{Chen}},
  \bibinfo{author}{\bibfnamefont{M.}~\bibnamefont{Hewitson}},
  \bibinfo{author}{\bibfnamefont{B.}~\bibnamefont{Krishnan}},
  \bibinfo{author}{\bibfnamefont{A.~M.} \bibnamefont{Sintes}},
  \bibinfo{author}{\bibfnamefont{J.~T.} \bibnamefont{Whelan}},
  \bibinfo{author}{\bibfnamefont{B.}~\bibnamefont{Br\"{u}gmann}},
  \bibinfo{author}{\bibfnamefont{P.}~\bibnamefont{Diener}},
  \bibinfo{author}{\bibfnamefont{N.}~\bibnamefont{Dorband}},
  \bibnamefont{et~al.}, \bibinfo{journal}{Phys.\ Rev.\ D}
  \textbf{\bibinfo{volume}{77}}, \bibinfo{eid}{104017} (\bibinfo{year}{2008}).

\bibitem[{\citenamefont{Buonanno
  et~al.}(2007{\natexlab{b}})\citenamefont{Buonanno, Pan, Baker, Centrella,
  Kelly, McWilliams, and van Meter}}]{Buonanno2007}
\bibinfo{author}{\bibfnamefont{A.}~\bibnamefont{Buonanno}},
  \bibinfo{author}{\bibfnamefont{Y.}~\bibnamefont{Pan}},
  \bibinfo{author}{\bibfnamefont{J.~G.} \bibnamefont{Baker}},
  \bibinfo{author}{\bibfnamefont{J.}~\bibnamefont{Centrella}},
  \bibinfo{author}{\bibfnamefont{B.~J.} \bibnamefont{Kelly}},
  \bibinfo{author}{\bibfnamefont{S.~T.} \bibnamefont{McWilliams}},
  \bibnamefont{and} \bibinfo{author}{\bibfnamefont{J.~R.} \bibnamefont{van
  Meter}}, \bibinfo{journal}{Phys.\ Rev.\ D} \textbf{\bibinfo{volume}{76}},
  \bibinfo{pages}{104049} (\bibinfo{year}{2007}{\natexlab{b}}).

\bibitem[{\citenamefont{Damour and Nagar}(2008)}]{Damour2007a}
\bibinfo{author}{\bibfnamefont{T.}~\bibnamefont{Damour}} \bibnamefont{and}
  \bibinfo{author}{\bibfnamefont{A.}~\bibnamefont{Nagar}},
  \bibinfo{journal}{Phys.\ Rev.\ D} \textbf{\bibinfo{volume}{77}},
  \bibinfo{eid}{024043} (\bibinfo{year}{2008}).

\bibitem[{\citenamefont{Damour et~al.}(2008)\citenamefont{Damour, Nagar,
  Dorband, Pollney, and Rezzolla}}]{DN2007b}
\bibinfo{author}{\bibfnamefont{T.}~\bibnamefont{Damour}},
  \bibinfo{author}{\bibfnamefont{A.}~\bibnamefont{Nagar}},
  \bibinfo{author}{\bibfnamefont{E.~N.} \bibnamefont{Dorband}},
  \bibinfo{author}{\bibfnamefont{D.}~\bibnamefont{Pollney}}, \bibnamefont{and}
  \bibinfo{author}{\bibfnamefont{L.}~\bibnamefont{Rezzolla}},
  \bibinfo{journal}{Phys.\ Rev.\ D} \textbf{\bibinfo{volume}{77}},
  \bibinfo{eid}{084017} (\bibinfo{year}{2008}).

\bibitem[{\citenamefont{{Damour}
  et~al.}(2008{\natexlab{a}})\citenamefont{{Damour}, {Nagar}, {Hannam}, {Husa},
  and {Br{\"u}gmann}}}]{DN2008}
\bibinfo{author}{\bibfnamefont{T.}~\bibnamefont{{Damour}}},
  \bibinfo{author}{\bibfnamefont{A.}~\bibnamefont{{Nagar}}},
  \bibinfo{author}{\bibfnamefont{M.}~\bibnamefont{{Hannam}}},
  \bibinfo{author}{\bibfnamefont{S.}~\bibnamefont{{Husa}}}, \bibnamefont{and}
  \bibinfo{author}{\bibfnamefont{B.}~\bibnamefont{{Br{\"u}gmann}}},
  \bibinfo{journal}{Phys.\ Rev.\ D} \textbf{\bibinfo{volume}{78}},
  \bibinfo{pages}{044039} (\bibinfo{year}{2008}{\natexlab{a}}).

\bibitem[{\citenamefont{Boyle et~al.}(2008)\citenamefont{Boyle, Buonanno,
  Kidder, Mrou\'e, Pan, Pfeiffer, and Scheel}}]{Boyle2008a}
\bibinfo{author}{\bibfnamefont{M.}~\bibnamefont{Boyle}},
  \bibinfo{author}{\bibfnamefont{A.}~\bibnamefont{Buonanno}},
  \bibinfo{author}{\bibfnamefont{L.~E.} \bibnamefont{Kidder}},
  \bibinfo{author}{\bibfnamefont{A.~H.} \bibnamefont{Mrou\'e}},
  \bibinfo{author}{\bibfnamefont{Y.}~\bibnamefont{Pan}},
  \bibinfo{author}{\bibfnamefont{H.~P.} \bibnamefont{Pfeiffer}},
  \bibnamefont{and} \bibinfo{author}{\bibfnamefont{M.~A.}
  \bibnamefont{Scheel}}, \bibinfo{journal}{Phys.\ Rev.\ D}
  \textbf{\bibinfo{volume}{78}}, \bibinfo{eid}{104020} (\bibinfo{year}{2008}).

\bibitem[{\citenamefont{Damour and Nagar}(2009)}]{Damour2009a}
\bibinfo{author}{\bibfnamefont{T.}~\bibnamefont{Damour}} \bibnamefont{and}
  \bibinfo{author}{\bibfnamefont{A.}~\bibnamefont{Nagar}},
  \bibinfo{journal}{Phys. Rev.} \textbf{\bibinfo{volume}{D79}},
  \bibinfo{pages}{081503} (\bibinfo{year}{2009}).

\bibitem[{\citenamefont{Buonanno et~al.}(2009)\citenamefont{Buonanno, Pan,
  Pfeiffer, Scheel, Buchman, and Kidder}}]{Buonanno:2009qa}
\bibinfo{author}{\bibfnamefont{A.}~\bibnamefont{Buonanno}},
  \bibinfo{author}{\bibfnamefont{Y.}~\bibnamefont{Pan}},
  \bibinfo{author}{\bibfnamefont{H.~P.} \bibnamefont{Pfeiffer}},
  \bibinfo{author}{\bibfnamefont{M.~A.} \bibnamefont{Scheel}},
  \bibinfo{author}{\bibfnamefont{L.~T.} \bibnamefont{Buchman}},
  \bibnamefont{and} \bibinfo{author}{\bibfnamefont{L.~E.}
  \bibnamefont{Kidder}}, \bibinfo{journal}{\prd} \textbf{\bibinfo{volume}{79}},
  \bibinfo{pages}{124028} (\bibinfo{year}{2009}).

\bibitem[{\citenamefont{Buonanno and Damour}(1999)}]{Buonanno99}
\bibinfo{author}{\bibfnamefont{A.}~\bibnamefont{Buonanno}} \bibnamefont{and}
  \bibinfo{author}{\bibfnamefont{T.}~\bibnamefont{Damour}},
  \bibinfo{journal}{Phys. Rev. D} \textbf{\bibinfo{volume}{59}},
  \bibinfo{pages}{084006} (\bibinfo{year}{1999}).

\bibitem[{\citenamefont{Buonanno and Damour}(2000)}]{Buonanno00}
\bibinfo{author}{\bibfnamefont{A.}~\bibnamefont{Buonanno}} \bibnamefont{and}
  \bibinfo{author}{\bibfnamefont{T.}~\bibnamefont{Damour}},
  \bibinfo{journal}{Phys.\ Rev.\ D} \textbf{\bibinfo{volume}{62}},
  \bibinfo{pages}{064015} (\bibinfo{year}{2000}).

\bibitem[{\citenamefont{Damour et~al.}(2000)\citenamefont{Damour, Jaranowski,
  and Sch\"afer}}]{2000PhRvD..62h4011D}
\bibinfo{author}{\bibfnamefont{T.}~\bibnamefont{Damour}},
  \bibinfo{author}{\bibfnamefont{P.}~\bibnamefont{Jaranowski}},
  \bibnamefont{and}
  \bibinfo{author}{\bibfnamefont{G.}~\bibnamefont{Sch\"afer}},
  \bibinfo{journal}{Phys.\ Rev.\ D} \textbf{\bibinfo{volume}{62}},
  \bibinfo{pages}{084011} (\bibinfo{year}{2000}).

\bibitem[{\citenamefont{Damour}(2001)}]{Damour01c}
\bibinfo{author}{\bibfnamefont{T.}~\bibnamefont{Damour}},
  \bibinfo{journal}{Phys.\ Rev.\ D} \textbf{\bibinfo{volume}{64}},
  \bibinfo{pages}{124013} (\bibinfo{year}{2001}).

\bibitem[{\citenamefont{{Damour}
  et~al.}(2008{\natexlab{b}})\citenamefont{{Damour}, {Jaranowski}, and
  {Sch{\"a}fer}}}]{DJS08}
\bibinfo{author}{\bibfnamefont{T.}~\bibnamefont{{Damour}}},
  \bibinfo{author}{\bibfnamefont{P.}~\bibnamefont{{Jaranowski}}},
  \bibnamefont{and}
  \bibinfo{author}{\bibfnamefont{G.}~\bibnamefont{{Sch{\"a}fer}}},
  \bibinfo{journal}{\prd} \textbf{\bibinfo{volume}{78}},
  \bibinfo{pages}{024009} (\bibinfo{year}{2008}{\natexlab{b}}).

\bibitem[{\citenamefont{{Ajith} et~al.}(2009)\citenamefont{{Ajith}, {Hannam},
  {Husa}, {Chen}, {Bruegmann}, {Dorband}, {Mueller}, {Ohme}, {Pollney},
  {Reisswig} et~al.}}]{Ajith:2009bn}
\bibinfo{author}{\bibfnamefont{P.}~\bibnamefont{{Ajith}}},
  \bibinfo{author}{\bibfnamefont{M.}~\bibnamefont{{Hannam}}},
  \bibinfo{author}{\bibfnamefont{S.}~\bibnamefont{{Husa}}},
  \bibinfo{author}{\bibfnamefont{Y.}~\bibnamefont{{Chen}}},
  \bibinfo{author}{\bibfnamefont{B.}~\bibnamefont{{Bruegmann}}},
  \bibinfo{author}{\bibfnamefont{N.}~\bibnamefont{{Dorband}}},
  \bibinfo{author}{\bibfnamefont{D.}~\bibnamefont{{Mueller}}},
  \bibinfo{author}{\bibfnamefont{F.}~\bibnamefont{{Ohme}}},
  \bibinfo{author}{\bibfnamefont{D.}~\bibnamefont{{Pollney}}},
  \bibinfo{author}{\bibfnamefont{C.}~\bibnamefont{{Reisswig}}},
  \bibnamefont{et~al.} (\bibinfo{year}{2009}), \eprint{arXiv:0909.2867}.

\bibitem[{\citenamefont{Buonanno et~al.}(2006)\citenamefont{Buonanno, Chen, and
  Damour}}]{Buonanno06}
\bibinfo{author}{\bibfnamefont{A.}~\bibnamefont{Buonanno}},
  \bibinfo{author}{\bibfnamefont{Y.}~\bibnamefont{Chen}}, \bibnamefont{and}
  \bibinfo{author}{\bibfnamefont{T.}~\bibnamefont{Damour}},
  \bibinfo{journal}{Phys.\ Rev.\ D} \textbf{\bibinfo{volume}{74}},
  \bibinfo{eid}{104005} (\bibinfo{year}{2006}).

\bibitem[{\citenamefont{Barausse et~al.}(2009)\citenamefont{Barausse, Racine,
  and Buonanno}}]{BRB}
\bibinfo{author}{\bibfnamefont{E.}~\bibnamefont{Barausse}},
  \bibinfo{author}{\bibfnamefont{E.}~\bibnamefont{Racine}}, \bibnamefont{and}
  \bibinfo{author}{\bibfnamefont{A.}~\bibnamefont{Buonanno}},
  \bibinfo{journal}{Phys. Rev.} \textbf{\bibinfo{volume}{D80}},
  \bibinfo{pages}{104025} (\bibinfo{year}{2009}).

\bibitem[{\citenamefont{{Pan} et~al.}(2009)\citenamefont{{Pan}, {Buonanno},
  {Fujita}, {Racine}, and {Tagoshi}}}]{Pan:2009}
\bibinfo{author}{\bibfnamefont{Y.}~\bibnamefont{{Pan}}},
  \bibinfo{author}{\bibfnamefont{A.}~\bibnamefont{{Buonanno}}},
  \bibinfo{author}{\bibfnamefont{R.}~\bibnamefont{{Fujita}}},
  \bibinfo{author}{\bibfnamefont{E.}~\bibnamefont{{Racine}}}, \bibnamefont{and}
  \bibinfo{author}{\bibfnamefont{H.}~\bibnamefont{{Tagoshi}}}
  (\bibinfo{year}{2009}), \bibinfo{note}{in preparation}.

\bibitem[{\citenamefont{Damour et~al.}(2009)\citenamefont{Damour, Iyer, and
  Nagar}}]{DIN}
\bibinfo{author}{\bibfnamefont{T.}~\bibnamefont{Damour}},
  \bibinfo{author}{\bibfnamefont{B.~R.} \bibnamefont{Iyer}}, \bibnamefont{and}
  \bibinfo{author}{\bibfnamefont{A.}~\bibnamefont{Nagar}},
  \bibinfo{journal}{Phys. Rev.} \textbf{\bibinfo{volume}{D79}},
  \bibinfo{pages}{064004} (\bibinfo{year}{2009}).

\bibitem[{\citenamefont{{Chu} et~al.}(2009)\citenamefont{{Chu}, {Pfeiffer}, and
  {Scheel}}}]{Chu:2009md}
\bibinfo{author}{\bibfnamefont{T.}~\bibnamefont{{Chu}}},
  \bibinfo{author}{\bibfnamefont{H.~P.} \bibnamefont{{Pfeiffer}}},
  \bibnamefont{and} \bibinfo{author}{\bibfnamefont{M.~A.}
  \bibnamefont{{Scheel}}} (\bibinfo{year}{2009}), \eprint{arXiv:0909.1313}.

\bibitem[{\citenamefont{Scheel et~al.}(2009)\citenamefont{Scheel, Boyle, Chu,
  Kidder, Matthews, and Pfeiffer}}]{Scheel2008}
\bibinfo{author}{\bibfnamefont{M.~A.} \bibnamefont{Scheel}},
  \bibinfo{author}{\bibfnamefont{M.}~\bibnamefont{Boyle}},
  \bibinfo{author}{\bibfnamefont{T.}~\bibnamefont{Chu}},
  \bibinfo{author}{\bibfnamefont{L.~E.} \bibnamefont{Kidder}},
  \bibinfo{author}{\bibfnamefont{K.~D.} \bibnamefont{Matthews}},
  \bibnamefont{and} \bibinfo{author}{\bibfnamefont{H.~P.}
  \bibnamefont{Pfeiffer}}, \bibinfo{journal}{Phys.\ Rev.\ D}
  \textbf{\bibinfo{volume}{79}}, \bibinfo{pages}{024003}
  (\bibinfo{year}{2009}).

\bibitem[{\citenamefont{Damour and Nagar}(2007)}]{Damour2007}
\bibinfo{author}{\bibfnamefont{T.}~\bibnamefont{Damour}} \bibnamefont{and}
  \bibinfo{author}{\bibfnamefont{A.}~\bibnamefont{Nagar}},
  \bibinfo{journal}{Phys.\ Rev.\ D} \textbf{\bibinfo{volume}{76}},
  \bibinfo{eid}{064028} (\bibinfo{year}{2007}).

\bibitem[{\citenamefont{Blanchet et~al.}(2006)\citenamefont{Blanchet, Buonanno,
  and Faye}}]{Blanchet-Buonanno-Faye:2006}
\bibinfo{author}{\bibfnamefont{L.}~\bibnamefont{Blanchet}},
  \bibinfo{author}{\bibfnamefont{A.}~\bibnamefont{Buonanno}}, \bibnamefont{and}
  \bibinfo{author}{\bibfnamefont{G.}~\bibnamefont{Faye}},
  \bibinfo{journal}{Phys.\ Rev.\ D} \textbf{\bibinfo{volume}{74}},
  \bibinfo{eid}{104034} (\bibinfo{year}{2006}).

\bibitem[{\citenamefont{Barack and Sago}(2009)}]{barack:2009}
\bibinfo{author}{\bibfnamefont{L.}~\bibnamefont{Barack}} \bibnamefont{and}
  \bibinfo{author}{\bibfnamefont{N.}~\bibnamefont{Sago}},
  \bibinfo{journal}{Phys.\ Rev.\ Lett.} \textbf{\bibinfo{volume}{102}},
  \bibinfo{eid}{191101} (\bibinfo{year}{2009}).

\bibitem[{\citenamefont{Damour}(2009)}]{Damour:2009sm}
\bibinfo{author}{\bibfnamefont{T.}~\bibnamefont{Damour}}
  (\bibinfo{year}{2009}), \eprint{arXiv:0910.5533}.

\bibitem[{\citenamefont{{Barausse} and {Buonanno}}(2009)}]{BB}
\bibinfo{author}{\bibfnamefont{E.}~\bibnamefont{{Barausse}}} \bibnamefont{and}
  \bibinfo{author}{\bibfnamefont{A.}~\bibnamefont{{Buonanno}}}
  (\bibinfo{year}{2009}), \bibinfo{note}{in preparation}.

\bibitem[{\citenamefont{Berti et~al.}(2009)\citenamefont{Berti, Cardoso, and
  Starinets}}]{Berti:2009}
\bibinfo{author}{\bibfnamefont{E.}~\bibnamefont{Berti}},
  \bibinfo{author}{\bibfnamefont{V.}~\bibnamefont{Cardoso}}, \bibnamefont{and}
  \bibinfo{author}{\bibfnamefont{A.~O.} \bibnamefont{Starinets}},
  \bibinfo{journal}{Class.\ Quantum Grav.} \textbf{\bibinfo{volume}{26}},
  \bibinfo{pages}{163001} (\bibinfo{year}{2009}).

\bibitem[{\citenamefont{Regge and Wheeler}(1957)}]{ReggeWheeler1957}
\bibinfo{author}{\bibfnamefont{T.}~\bibnamefont{Regge}} \bibnamefont{and}
  \bibinfo{author}{\bibfnamefont{J.~A.} \bibnamefont{Wheeler}},
  \bibinfo{journal}{Phys.\ Rev.} \textbf{\bibinfo{volume}{108}},
  \bibinfo{pages}{1063} (\bibinfo{year}{1957}).

\bibitem[{\citenamefont{Zerilli}(1970)}]{Zerilli1970b}
\bibinfo{author}{\bibfnamefont{F.~J.} \bibnamefont{Zerilli}},
  \bibinfo{journal}{Phys.\ Rev.\ Lett.} \textbf{\bibinfo{volume}{24}},
  \bibinfo{pages}{737} (\bibinfo{year}{1970}).

\bibitem[{\citenamefont{Sarbach and Tiglio}(2001)}]{Sarbach2001}
\bibinfo{author}{\bibfnamefont{O.}~\bibnamefont{Sarbach}} \bibnamefont{and}
  \bibinfo{author}{\bibfnamefont{M.}~\bibnamefont{Tiglio}},
  \bibinfo{journal}{Phys. Rev. D} \textbf{\bibinfo{volume}{64}},
  \bibinfo{pages}{084016} (\bibinfo{year}{2001}).

\bibitem[{\citenamefont{{Rinne} et~al.}(2009)\citenamefont{{Rinne}, {Buchman},
  {Scheel}, and {Pfeiffer}}}]{Rinne2008b}
\bibinfo{author}{\bibfnamefont{O.}~\bibnamefont{{Rinne}}},
  \bibinfo{author}{\bibfnamefont{L.~T.} \bibnamefont{{Buchman}}},
  \bibinfo{author}{\bibfnamefont{M.~A.} \bibnamefont{{Scheel}}},
  \bibnamefont{and} \bibinfo{author}{\bibfnamefont{H.~P.}
  \bibnamefont{{Pfeiffer}}}, \bibinfo{journal}{Class.\ Quantum Grav.}
  \textbf{\bibinfo{volume}{26}}, \bibinfo{pages}{075009}
  (\bibinfo{year}{2009}).

\bibitem[{\citenamefont{Pan et~al.}(2004)\citenamefont{Pan, Buonanno, Chen, and
  Vallisneri}}]{Pan2004}
\bibinfo{author}{\bibfnamefont{Y.}~\bibnamefont{Pan}},
  \bibinfo{author}{\bibfnamefont{A.}~\bibnamefont{Buonanno}},
  \bibinfo{author}{\bibfnamefont{Y.}~\bibnamefont{Chen}}, \bibnamefont{and}
  \bibinfo{author}{\bibfnamefont{M.}~\bibnamefont{Vallisneri}},
  \bibinfo{journal}{Phys.\ Rev.\ D} \textbf{\bibinfo{volume}{69}},
  \bibinfo{pages}{104017} (\bibinfo{year}{2004}).

\bibitem[{\citenamefont{Kidder}(2008)}]{Kidder:2007rt}
\bibinfo{author}{\bibfnamefont{L.~E.} \bibnamefont{Kidder}},
  \bibinfo{journal}{Phys.\ Rev.\ D} \textbf{\bibinfo{volume}{77}},
  \bibinfo{pages}{044016} (\bibinfo{year}{2008}).

\bibitem[{\citenamefont{Lindblom et~al.}(2008)\citenamefont{Lindblom, Owen, and
  Brown}}]{Lindblom2008}
\bibinfo{author}{\bibfnamefont{L.}~\bibnamefont{Lindblom}},
  \bibinfo{author}{\bibfnamefont{B.~J.} \bibnamefont{Owen}}, \bibnamefont{and}
  \bibinfo{author}{\bibfnamefont{D.~A.} \bibnamefont{Brown}},
  \bibinfo{journal}{\prd} \textbf{\bibinfo{volume}{78}},
  \bibinfo{pages}{124020} (\bibinfo{year}{2008}).

\bibitem[{\citenamefont{Boyle et~al.}(2007)\citenamefont{Boyle, Brown, Kidder,
  Mrou{\'e}, Pfeiffer, Scheel, Cook, and Teukolsky}}]{Boyle2007}
\bibinfo{author}{\bibfnamefont{M.}~\bibnamefont{Boyle}},
  \bibinfo{author}{\bibfnamefont{D.~A.} \bibnamefont{Brown}},
  \bibinfo{author}{\bibfnamefont{L.~E.} \bibnamefont{Kidder}},
  \bibinfo{author}{\bibfnamefont{A.~H.} \bibnamefont{Mrou{\'e}}},
  \bibinfo{author}{\bibfnamefont{H.~P.} \bibnamefont{Pfeiffer}},
  \bibinfo{author}{\bibfnamefont{M.~A.} \bibnamefont{Scheel}},
  \bibinfo{author}{\bibfnamefont{G.~B.} \bibnamefont{Cook}}, \bibnamefont{and}
  \bibinfo{author}{\bibfnamefont{S.~A.} \bibnamefont{Teukolsky}},
  \bibinfo{journal}{Phys.\ Rev.\ D} \textbf{\bibinfo{volume}{76}},
  \bibinfo{eid}{124038} (\bibinfo{year}{2007}).

\end{thebibliography}
%\end{document}

\end{document}